\documentclass[11pt, a4paper]{article}

\usepackage{geometry}
\usepackage{graphicx}
\usepackage{multirow}
\usepackage{amsmath,amssymb,amsfonts}
\usepackage{amsthm}
\usepackage{subcaption}
\usepackage{mathrsfs}
\usepackage[utf8]{inputenc}
\usepackage{textcomp}
\usepackage[title]{appendix}
\usepackage{xcolor}
\usepackage{textcomp}
\usepackage{manyfoot}
\usepackage{booktabs}
\usepackage{algorithm}
\usepackage{algorithmicx}
\usepackage{orcidlink}
\usepackage{algpseudocode}
\usepackage{float}
\usepackage{listings}
\lstdefinestyle{circuitstyle}{
  language=Python,
  basicstyle=\ttfamily\small,
  keywordstyle=\color{black},
  breaklines=true,
}
\usepackage{authblk}
\usepackage{braket}
\usepackage{physics}
\usepackage{bbm}
\usepackage{cite}
\usepackage{tcolorbox}
\tcbuselibrary{theorems}

\theoremstyle{plain}

\theoremstyle{definition}

\newtcbtheorem[number within=section]{prompt}{Prompt}%
{colback=green!5,colframe=green!35!black,fonttitle=\bfseries}{th}

\begin{document}

\title{\textbf{Fine-Tuning Large Language Models for Quantum Reasoning}}

\author[1]{Katherine Ip\thanks{Email:katherine.ip@student.unimelb.edu.au}\orcidlink{0009-0006-9042-1731}}
\author[2,3]{Casey R. Myers\orcidlink{0000-0002-8838-7523}}
\author[1]{Udaya Parampalli\orcidlink{0000-0002-9798-0134}}
\author[4]{James Quach\orcidlink{0000-0002-3619-2505}}
\author[4]{Peiyong Wang\orcidlink{0000-0002-0665-6639}}

\affil[1]{School of Computing and Information Systems, The University of Melbourne, Melbourne, Australia}
\affil[2]{School of Physics, Mathematics and Computing, The University of Western Australia, 35 Stirling Hwy, Crawley WA, 6009, Australia}
\affil[3]{Pawsey Supercomputing Centre, 1 Bryce Avenue, Kensington WA, 6151, Australia}
\affil[4]{CSIRO Clayton, Research Way, Clayton VIC 3168, Australia}

\date{}

\maketitle

\begin{abstract}
Large language models (LLMs) exhibit abilities beyond natural language modelling and text generation. Recent advances in their reasoning capabilities have spurred interest in applying LLMs to complex scientific tasks requiring deep domain expertise and sophisticated reasoning. Quantum computing, as a highly specialised field with significant knowledge barriers and hardware constraints, could greatly benefit from such advancements. However, a key open question that first must be answered is: \textit{How can we develop fine-tuning pipelines that instil genuine quantum reasoning in LLMs, rather than task-specific pattern matching?} We study this question through \textit{quantum circuit simulation} as a training objective, where the model must predict the measurement probability distribution resulting from a sequence of quantum gate operations. We propose and compare two fine-tuning pipelines: (1) Supervised Fine-Tuning (SFT) on explicit gate-by-gate state-vector simulation traces, and (2) a two-stage SFT+Group Relative Policy Optimisation (GRPO) approach that sequentially applies SFT followed by GRPO with verifiable rewards. Our findings show that SFT achieves near-perfect in-distribution and gate-count extrapolation accuracy, significantly outperforming both the base model and the GPT-OSS-120B baseline. SFT+GRPO trades some in-distribution precision for better generalisation to larger qubit systems that SFT alone cannot handle. Both pipelines significantly outperform the baselines, demonstrating that targeted fine-tuning on explicit reasoning traces is an effective strategy for advancing quantum reasoning in LLMs.
\end{abstract}

\textbf{Keywords:} Large Language Models (LLM), Quantum Computing, LLM Fine-Tuning, LLM Reasoning

\section{Introduction}\label{sec:intro}

Large language models (LLMs) are built as next-token predictors, but recent models exhibit abilities that exceed simple word prediction \cite{gandhi2025cognitive, qwen2025qwen25technicalreport, dubey2024llama}. This observation raises questions about the nature of their reasoning, which can be further explored through Dual Process Theory \cite{kahneman2011thinking}. The theory proposes two systems of cognitive reasoning: the automated, intuitive "System 1" fast reasoning, and the deliberative, complex "System 2" slow reasoning. Researchers are currently pushing LLMs toward System 2 reasoning, demonstrating LLMs’ ability to tackle advanced mathematics \cite{chervonyi2025gold}, code reasoning and generation \cite{carbonneaux2025cwm}, and algorithm design \cite{liu2024systematic}, with the goal of moving toward artificial general intelligence (AGI) \cite{bubeck2023sparks}.

Solving challenges in scientific domains requires System 2 thinking and domain expertise. LLMs are pretrained on vast text corpora to encode deep knowledge from scientific domains. With recent improvements in LLMs' reasoning capabilities, they show potential to assist scientific discovery. For example, Zhong et al. \cite{zhong2024benchmarking} explore LLMs in predicting molecular properties through prompt engineering. In another application, Ramos et al. \cite{ramos2025bayesianoptimizationcatalysisincontext} use LLMs as surrogate models for Bayesian optimisation in catalyst design. More ambitiously, Yamada et al. \cite{yamada2025ai} built an end-to-end LLM agentic system to automate the full machine learning research lifecycle and produce workshop-level research papers.  

As in other scientific domains, there is growing interest in the quantum computing community in using LLMs to support research and development. Early applications have  on computing aspects, where LLMs leverage their code understanding and generation capabilities \cite{jiang2024survey} and serve as \textit{quantum coding agents} performing tasks including quantum program repair~\cite{guo2024repairing}, code-to-code translation from classical to quantum machine learning~\cite{zeng2026q}, and code generation for Qiskit~\cite{dupuis2024qiskit, dupuis2025quantum, campbell2025enhancing, mikuriya2025qcoder}, PennyLane~\cite{basit2025pennylang, basit2025pennycoder}, as well as domain-specific programming language OpenQASM programming~\cite{fu2025qagent}.

More recently, researchers have shifted from quantum \textit{code} generation to quantum \textit{circuit} generation, where LLMs have been used to design circuits that solve computational challenges. Recent work has focussed on variational quantum circuits: Liang et al.~\cite{liang2023unleashing} proposed a multi-agent framework to generate parameterised circuits from natural-language problem descriptions; Jern et al.~\cite{jern2025fine}, and Yu et al.~\cite{yu2025quasar} trained LLMs to generate and initialise QAOA and VQE circuits for combinatorial optimisation; and Hive~\cite{finger2026automated} and IdeaSearch~\cite{cao2025scalable} incorporated LLMs to discover compact VQE-style ans\"{a}tze for molecular simulation and quantum field theory. LLM-guided agentic systems have also been applied to quantum machine learning (QML) circuit design, including ans\"{a}tz search for quantum generative models~\cite{ueda2025optimizing} and automated feature map discovery~\cite{sakka2025automating}. This shift in focus from computing code generation to a more central quantum perspective demands a higher level of domain knowledge of quantum information and quantum computation theory. Beyond circuit generation, GroverGPT~\cite{wang2024grovergpt} and its successor GroverGPT+~\cite{chen2026symbolic} pushed further by fine-tuning LLMs to simulate and symbolically analyse Grover’s search quantum circuits.

Prior work has demonstrated practical applications of LLMs in this domain, either relying on prompt engineering and agentic frameworks that leverage pretrained quantum knowledge without task-specific training, or applying fine-tuning narrowly to specific tasks, such as quantum code generation or circuit synthesis, for fixed problem types. Neither of these approaches directly targets general quantum reasoning, defined as the ability to reason about arbitrary quantum operations and generalise to unseen circuits. The gap, therefore, lies in developing fine-tuning pipelines that instil better quantum reasoning rather than task-specific pattern matching.

In this work, we address this gap through the lens of \textit{quantum circuit simulation}: given a quantum circuit described as a sequence of gate operations, our goal is to train an LLM to reason through the gates and predict the resulting measurement probability distribution. This framing is inspired by the Code World Model (CWM)~\cite{carbonneaux2025cwm}, which demonstrates that training LLMs to predict program execution traces can substantially improve their code-reasoning capabilities and, in turn, overall performance in code generation. By analogy, we conjecture that training LLMs on explicit gate-by-gate state-vector simulation traces can ground the model in quantum mechanical principles, providing a principled training signal for strengthening quantum reasoning. We make the following contributions:

\begin{enumerate}
    \item \sloppy We propose and evaluate two fine-tuning strategies for LLM quantum reasoning on non-parameterised and parameterised quantum circuits spanning 1--5 qubits and 1--50 gates: (i) SFT on explicit gate-by-gate state-vector simulation traces, which achieves the highest in-distribution accuracy and token efficiency with robust gate-count extrapolation; and (ii)~a two-stage SFT+GRPO pipeline that first instils structured quantum reasoning via SFT, then applies GRPO with verifiable rewards to encourage exploration of more efficient reasoning strategies, resulting in partial generalisation to larger qubit systems where SFT collapses entirely. Both pipelines significantly outperform the base model Qwen3-8B and the GPT-OSS-120B baseline.
    \item \sloppy We introduce step-by-step state fidelity as an evaluation metric for the state-vector simulation traces. This metric computes the quantum state fidelity at each reasoning step, going beyond final-answer accuracy by localising the step at which the model’s quantum reasoning degrades. This approach allows for better interpretability by enabling investigation into where and how LLM quantum reasoning deteriorates over the course of the steps.
    \item Through an ablation study comparing SFT with and without the explicit gate-by-gate simulation traces, we show that LLMs can approximate quantum circuit measurement outcomes in the latent space. LLMs trained without reasoning traces still predict with moderate accuracy but exhibit a higher total variation distance (TVD) than SFT with reasoning traces. This result demonstrates that step-by-step reasoning traces provide an additional learning signal that helps LLMs reason about quantum circuits more accurately.
    \item We show that length generalisation is the primary bottleneck when applying LLMs to quantum computing tasks. The results of our qualitative analysis show that both models initialise an incorrectly-sized state vector for out-of-distribution qubit counts, though the SFT+GRPO models partially recover at the output level by producing correctly-sized bitstrings where the SFT models do not.

\end{enumerate}

The remainder of this paper is organised as follows. Section~\ref{sec:prelim} introduces the technical background on transformers, supervised fine-tuning, and reinforcement learning with GRPO. Section~\ref{sec:related} surveys related work on LLMs for quantum computing and quantum circuit simulation. Section~\ref{sec:method} describes our circuit datasets and the two training pipelines (SFT and SFT+GRPO). Section~\ref{sec:results} presents results across in-distribution, gate-count extrapolation, and system-size extrapolation settings. Section~\ref{sec:discussion} discusses the strengths and limitations of the two pipelines and concludes with directions for future work.

\section{Preliminaries}\label{sec:prelim}

\subsection{Transformer}

LLMs are built upon the transformer architecture \cite{vaswani2017attention}, which models sequences using the self-attention mechanism. The model operates on sequences of \textit{tokens}, which can represent words, subwords, or characters depending on the tokenisation algorithm.  While the original transformer employs an encoder-decoder architecture, where the encoder produces latent representations of the input and the decoder autoregressively generates the output, most LLMs now rely solely on the decoder stack for autoregressive text generation \cite{radford2018improving}. In decoder-only models parameterised by $\boldsymbol{\theta}$, a given sequence of tokens $x = (x_1, x_2,...,x_T)$ , including both input and previously generated tokens, is treated uniformly as a prefix. The model performs causal (left-to-right) next-token prediction by calculating the conditional probability of the next token $x_{t+1}$ given all preceding tokens:

\begin{equation}
p_{\boldsymbol{\theta}}(x_{t+1} | x_1, \ldots, x_t).
\end{equation}

The self-attention mechanism of transformers computes contextualised word representations by allowing each token to weigh the importance of all other tokens in the input sequence. For each token, the mechanism first transforms it into three vectors---\textit{Query ($Q$), Key ($K$), and Value ($V$)}---through learned linear projections \cite{vaswani2017attention}. The attention operation then computes weighted combinations of values based on attention scores calculated from these vectors:

\begin{equation}
\text{Attention}(Q, K, V) = \text{softmax}\left(\frac{Q K^T}{\sqrt{d_k}}\right) V,
\end{equation}
where $d_k$ is the dimension of the key vectors, and the scaling factor $1/\sqrt{d_k}$ prevents the dot products from growing too large in magnitude. The softmax function, defined as 

\begin{equation}
\text{softmax}(x_i) = \frac{e^{x_i}}{\sum_{j} e^{x_j}},
\end{equation}
where $x_i$ represents each element of the scaled dot-product matrix, normalises the attention weights to ensure they sum to one across each query.

Transformers use \textit{multi-head attention} that projects the queries, keys, and values multiple times with different learned projections. For multi-head attention with $h$ heads:

\begin{equation}
\text{MultiHead}(Q, K, V) = \text{Concat}(\text{head}_1, \ldots, \text{head}_h) W^O, 
\end{equation}
where $\text{head}_i = \text{Attention}(Q W_i^Q, K W_i^K, V W_i^V)$, with $W_i^Q \in \mathbb{R}^{d_{\text{model}} \times d_k}, \quad
W_i^K \in \mathbb{R}^{d_{\text{model}} \times d_k}, \quad
W_i^V \in \mathbb{R}^{d_{\text{model}} \times d_v}$ are parameter matrices and $W^O$ is an output projection matrix. This mechanism allows the model to simultaneously process information from different representation subspaces at different positions, thereby capturing diverse features in a sequence.

\subsection{Supervised Fine-Tuning (SFT)}
Formally, given a dataset $\mathcal{D} = \{(x_i, y_i)\}_{i=1}^N$ of $N$ input-output pairs, SFT is formulated as a sequence of next-token prediction tasks \cite{radford2018improving}. For a sequence with $T$ tokens, the model minimises the negative log-likelihood as follows:

\begin{equation}
\mathcal{L}_{\text{SFT}}(\boldsymbol{\theta}) = -\sum_{t=1}^{T} \log p_{\boldsymbol{\theta}} (y_t \mid x, y_{<t}),
\end{equation}
where $y_t$ denotes the target token at timestep $t$, $y_{<t}$ represents all preceding tokens. To ensure the model learns to generate the response rather than reconstruct the input, the loss on the prompt tokens $x$ is masked, focussing the gradient updates solely on the response tokens $y$.

\subsection{Reinforcement Learning (RL)}
The generation process of an LLM can be formulated as a Markov Decision Process (MDP) \cite{pmlr-v235-wan24c}, defined by the tuple $(\mathcal{S}, \mathcal{A}, \mathcal{R}, \mathcal{T}, \pi_\theta)$:

\begin{itemize}
    \item \textbf{States $\mathcal{S}$:} A state $s_t \in \mathcal{S}$ consists of the initial prompt $x$ and the sequence of tokens generated up to step $t$, such that $s_t = (x, y_{<t})$. The initial state is defined as $s_0 = x$, while the terminal state $s_T$ represents the complete sequence $(x, y_1, \dots, y_T)$.

\item \textbf{Actions $ \mathcal{A}$:} The action space $A$ corresponds to the LLM’s vocabulary $\mathcal{V}$, which is the set of all possible tokens. An action $a_t \in A$ at step $t$ corresponds to selecting the next token $y_t$ from $\mathcal{V}$ for the sequence.

\item \textbf{Transition Function $\mathcal{T}$:} Given the current state $s_t$ and the chosen next token $a_t$, the next state is determined by $s_{t+1} = \mathcal{T}(s_t, a_t)$. In the context of LLM fine-tuning, the next state is the concatenation of sequence $s_t$ with token $a_t$.

\item \textbf{Reward Function $\mathcal{R}$:} The reward function assigns a scalar value $r_t = R(s_t, a_t)$ that indicates the quality of the generated sequence. For LLM fine-tuning, rewards are often sparse, with only terminal states $s_T$ receiving non-zero rewards.

\item \textbf{Policy $\pi_{\boldsymbol{\theta}}$:} The policy $\pi_{\boldsymbol{\theta}}$ is the LLM. The parameters $\boldsymbol{\theta}$ define the probability distribution $\pi_{\boldsymbol{\theta}} (a_t|s_t)$ over possible next tokens $a_t$ given the current sequence $s_t$.
\end{itemize}

\subsubsection{Group Relative Policy Optimisation (GRPO)}
The fine-tuning process with reinforcement learning is framed as a policy optimisation problem, with the objective of optimising policy parameters $\theta$ to maximise the expected cumulative reward \cite{guo2025deepseek}.

GRPO \cite{shao2024deepseekmath} is an on-policy RL algorithm for LLM fine-tuning that eliminates the need for computationally expensive value function approximations by replacing them with the mean reward from a group of sampled outputs.

\sloppy For each input-output pair $(x, y)$ from data distribution $\mathcal{D}$, $G$ responses $\{o_1, o_2,\ldots, o_G\}$ are sampled from the current policy $\pi_{\theta_\text{old}}$. Each response $o_i$ consists of a sequence of tokens $o_i = (o_{i,1}, o_{i,2}, \ldots, o_{i,|o_i|})$, where $o_{i,t}$ denotes the $t$-th token. A reward $r_i$ is computed for each sample, forming the reward vector $\mathbf{r} = \{r_1, r_2,\ldots, r_G\}$. The model is optimised by maximising the following objective:

\begin{equation}
\mathcal{J}_\text{GRPO} = \mathbb{E}_{(q,a) \sim \mathcal{D}, \{o_i\}^G_{i = 1} \sim \pi_{\theta_{\text{old}}}(\cdot|q)} [\mathcal{L}_\text{clip}(\theta) - \beta D_{\text{KL}}[\pi_\theta||\pi_\text{ref}]], 
\end{equation}
where the clipped loss is:

\begin{equation}
\mathcal{L}_\text{clip}(\theta) = \frac{1}{G} \sum^G_{i = 1} \frac{1}{|o_i|} \sum^{|o_i|}_{t = 1} \min(\rho_{i,t} \hat{A}_{i,t}, \text{clip}(\rho_{i,t}, 1 - \epsilon, 1 + \epsilon) \hat{A}_{i,t})
\end{equation}
and 
\begin{equation}
\rho_{i,t} = \frac{\pi_\theta (o_{i,t} | q, o_{i, <t})}{\pi_{\theta_\text{old}} (o_{i,t} | q, o_{i, <t})}
\end{equation}
in the clipped loss is the probability ratio between the current policy and the old policy, and $\text{clip}(\rho_{i,t}, 1 - \epsilon, 1 + \epsilon)$ constrains this ratio to the interval $[1-\epsilon, 1+\epsilon]$ to prevent excessively large policy updates. The parameters $\pi_\theta$ and $\pi_\text{ref}$ represent the updated model and the reference model (i.e., the original model before fine-tuning), respectively, and $\beta$ and $\epsilon$ are hyperparameters.
The advantage $\hat{A}_{i,t}$ is computed as the normalised relative comparison:
\begin{equation}
\hat{A}_{i,t} = \frac{r_i - \text{mean}(\mathbf{r})}{\text{std}(\mathbf{r})}.
\end{equation}
The KL divergence $D_{\text{KL}}$ is computed as:
\begin{equation}
D_{\text{KL}}[\pi_\theta||\pi_\text{ref}] = \frac{\pi_\text{ref} (o_{i,t} | q, o_{i, <t})}{\pi_\theta (o_{i,t} | q, o_{i, <t})} - \log \frac{\pi_\text{ref} (o_{i,t} | q, o_{i, <t})}{\pi_\theta (o_{i,t} | q, o_{i, <t})} - 1.
\end{equation}

\subsubsection{Reinforcement Learning with Verifiable Reward (RLVR)}
Unlike SFT, fine-tuning with reinforcement learning does not require ground truth; instead, a reward is provided for each generated output to encourage the model to explore desirable outputs. RLVR \cite{lambert2024t} is one of the RL fine-tuning approaches where reward signals are computed by an external verifier. This approach is effective at preventing reward hacking and at incentivising the model to explore correct reasoning traces that lead to correct answers. For example, DeepSeek-R1 \cite{guo2025deepseek} uses a rule-based reward system for verifiable domains such as mathematics, coding, and logical reasoning. The verifier computes two binary rewards: accuracy and format. The model receives a format reward when the output adheres to specific structural constraints, and an accuracy reward when the final answer matches the ground truth. The reasoning traces are not being evaluated, which encourages the model to freely explore effective reasoning traces that increase the accuracy of the final prediction. The result shows that, with RLVR, models autonomously developed complex reasoning behaviours, such as self-verification and reflection, and achieved superior performance on complex reasoning tasks compared to counterparts trained solely with SFT.

\subsection{Low-Rank Adaptation (LoRA)}
As LLMs grow in size, fully updating all parameters during fine-tuning becomes 
computationally prohibitive. Consider a weight matrix $W_0 \in \mathbb{R}^{d \times k}$, 
where $d$ is the embedding dimension and $k$ is the output dimension. Full-parameter 
fine-tuning requires computing and storing a weight update matrix of the same size as 
$W_0$, which is infeasible at scale. LoRA \cite{hu2022lora} addresses this by 
approximating the weight update $\Delta W$ with a low-rank decomposition $BA$, where 
$B \in \mathbb{R}^{d \times r}$, $A \in \mathbb{R}^{r \times k}$, and $r \ll \min(d, k)$, 
reducing the number of trainable parameters from $d \times k$ to $r(d + k)$. During 
fine-tuning, $W_0$ is frozen and only $A$ and $B$ are trained. The update is scaled by 
$\alpha / r$, where $\alpha$ is a hyperparameter controlling the magnitude of the 
adaptation, so the modified forward pass becomes:
\begin{equation}
    h = W_0 x + \frac{\alpha}{r} B A x.
\end{equation}
A higher rank $r$ retains more expressive capacity at the cost of more parameters and 
longer training, while a lower rank reduces cost but risks losing task-relevant information. 
Despite the substantial reduction in trainable parameters, LoRA achieves performance 
comparable to full-parameter fine-tuning across various tasks \cite{hu2022lora}.

\section{Related Work}\label{sec:related}

\subsection{LLM for Quantum Computing}

An emerging direction leverages prompt engineering and multi-agent systems to apply the quantum knowledge encoded in pretrained LLMs without any task-specific training. For quantum coding tasks, retrieval-augmented generation (RAG) over quantum library documentation has been shown to improve PennyLane code generation~\cite{basit2025pennylang}. QAgent~\cite{fu2025qagent} pairs a RAG-driven dynamic few-shot coder with a tools-augmented coder that invokes a predefined quantum programming library, with a reflection agent revising outputs against simulator feedback, achieving a 71.6\% improvement over static few-shot baselines on OpenQASM generation. QCoder~\cite{mikuriya2025qcoder} similarly feeds structured simulator error reports back to the model as natural language for iterative refinement, raising o3's pass rate from 65.5\% to approximately 78\%. Sakka et al.~\cite{sakka2025automating} take this further with a five-component closed loop that generates, validates, evaluates, and reviews quantum feature maps using classification accuracy as the reward signal, autonomously discovering circuits that outperform all tested quantum baselines on a classification benchmark MNIST. At a larger scale, Hive~\cite{finger2026automated} and IdeaSearch~\cite{cao2025scalable} embed LLMs inside evolutionary algorithm search loops where the model proposes and mutates ansatz circuit structures, with a sandboxed evaluator scoring each candidate on energy deviation and circuit resource metrics, yielding VQE-style ans\"{a}tze that achieve chemical precision with orders-of-magnitude fewer circuit evaluations than ADAPT-VQE.

Alternatively, other work applied fine-tuning to directly enhance LLMs' quantum knowledge. Most of the work utilises supervised fine-tuning (SFT) on quantum-specific data. For example, Dupuis et al.~\cite{dupuis2024qiskit} apply extended pretraining of a 20-billion (20B)-parameter Granite code model on 88-million tokens of Qiskit scripts and Jupyter notebooks scraped from GitHub, followed by instruction tuning on synthetic Qiskit question-answer pairs. This raised pass@1 on the Qiskit HumanEval benchmark from 20.79\% to 46.53\%, outperforming larger general-purpose code models. PennyCoder~\cite{basit2025pennycoder} applied LoRA fine-tuning to LLaMA 3.1-8B on a curated PennyLane instruction dataset, outperforming both the base model and retrieval-augmented generation (RAG) on a held-out test set of PennyLane code generation problems. Campbell et al.~\cite{campbell2025enhancing} provided further evidence by showing that Chain-of-Thought (CoT) and structured CoT (SCoT) elicit LLM's quantum knowledge more effectively than RAG. For circuit generation, Jern et al.~\cite{jern2025fine} fine-tuned an LLM on 14,000 parameterised QAOA and VQE circuits using SFT, producing syntactically correct OpenQASM circuits with near-optimal initial parameter values that outperformed random initialisation on both expectation value and output distribution metrics.

More recent work has added an additional GRPO stage after SFT, using verifiable reward signals grounded in quantum execution. Dupuis et al.~\cite{dupuis2025quantum} built a pipeline that first distils the SFT model using DeepSeek-V3 as a teacher, then applied Direct Preference Optimisation (DPO) and GRPO, with the reward defined as the unit-test pass rate when the generated code is executed on a quantum simulator. Comparing DPO and GRPO directly, they found that GRPO specifically learned to self-correct module import errors through self-play exploration, a behaviour absent in the DPO-trained counterpart and also outperformed off-the-shelf model with more parameters. QUASAR~\cite{yu2025quasar} applied GRPO to OpenQASM circuit generation. QUASAR initialised from Jern et al.'s SFT model~\cite{jern2025fine} and used a four-level hierarchical reward progressed from syntactic validity, to output-distribution fidelity, to Hamiltonian expectation-value alignment, and finally to optimisation convergence speed. Despite using only a 4B model, QUASAR outperformed GPT-5 and GPT-4o on all metrics. Both works initialised GRPO from an SFT model rather than the base model, following the cold-start strategy of DeepSeek-R1~\cite{guo2025deepseek}, and both targeted quantum \textit{coding} tasks where correctness was defined by execution against a known library or Hamiltonian.

While circuit synthesis demonstrated that LLMs can generate quantum circuits, a fundamental question remains: \textit{Do these results reflect a genuine understanding of quantum computing theory, or do they primarily reflect pattern recognition on familiar circuit structures from training data?} Similar to observations in LLM code generation research, where producing correct code does not necessarily imply deep code understanding \cite{haroon2025accurately}, the ability to generate quantum circuits does not guarantee that LLMs engage in actual quantum reasoning. Without a genuine understanding of quantum mechanics, LLM-generated circuits may rely on superficial patterns learned from training data rather than principled quantum reasoning, limiting their ability to generalise to novel problems and remain robust to variations in problem specifications. 

\subsection{Transformer Models for Quantum Circuit Simulation and Symbolic Analysis}

Quantum circuit simulation involves computing or approximating the output quantum state resulting from applying a quantum circuit to a specific initial state with classical computational resources. This task requires understanding how quantum gates manipulate qubit states and how these operations compose in a circuit. Conventional methods include brute-force state-vector simulation and tensor network approximation methods \cite{orus2014practical}. More recently, machine learning approaches have introduced neural quantum states that learn compressed representations of quantum states \cite{carleo2017solving}, offering greater flexibility in capturing complex entanglement patterns.

Building on the neural representation approach, Zhou et al. \cite{zhou2025application} investigated whether a transformer model can learn to predict quantum-state evolution from circuit parameters. They formulated quantum simulation as a regression problem: given circuit parameters as input, predict the amplitudes of the output quantum state vector. For single-qubit systems, they parameterised arbitrary quantum gates using the universal single-qubit gate $U(\theta, \phi, \lambda)$, which can generate any single-qubit quantum state. The model takes three rotation angles as input and predicts the four components of the output state vector: two real parts ($a$, $c$) and two imaginary parts ($b$, $d$), representing the complex state $\begin{pmatrix} a + bi \\ c + di \end{pmatrix}$. Using an encoder-decoder transformer architecture fine-tuned on gate-parameter and state-value pairs, the model achieved over 99\% fidelity in predicting single-qubit state evolution. The study further extended this approach to systems with two and three qubits, achieving fidelities above 94.7\% and approximately 93\%, respectively. These findings indicate that transformer architectures can learn the mapping from circuit parameters to quantum states for small-scale quantum systems, and demonstrate the potential of LLMs to learn quantum state evolution and handle quantum circuit simulation tasks.

A more recent work, GroverGPT \cite{wang2024grovergpt}, explored LLMs' ability to perform quantum circuit simulation directly from textual circuit descriptions. Unlike Zhou et al.'s encoder-decoder architecture trained from scratch, GroverGPT employed a pretrained transformer, Llama-3.1-8B \cite{dubey2024llama}, leveraging the model's pretrained knowledge and language understanding to potentially capture the quantum transformations more effectively. This work was fine-tuned on 97K examples of Grover's search algorithm circuits \cite{grover1996fast} with their corresponding output probability distributions. However, GroverGPT's simulation scope is limited: the model was trained and evaluated solely on optimised Grover's search algorithm circuits, which have a specific, repetitive gate structure. This constrained the claim as a general quantum circuit simulator and raises the question of whether the model captures genuine quantum reasoning or exploits structural patterns specific to Grover's algorithm.

Its successor, GroverGPT+ \cite{chen2026symbolic}, reframed the problem by shifting from numerical simulation to symbolic analysis.  GroverGPT+ aimed to interpret the high-level algorithmic structure of a quantum circuit and articulate its logic in a human-readable format. Given a QASM description of a Grover circuit, GroverGPT+ generated an interpretable reasoning trace that identified the oracle and the corresponding marked states, then inferred the output probability distribution from this symbolic understanding. Evaluated on circuits ranging from 2 to 9 qubits, the model achieved search accuracy and classical fidelity consistently approaching 1.0 within its training distribution. However, it is unclear whether this strong performance was due to genuine algorithmic understanding or to superficial pattern matching. The format of the QASM circuits used in the study was specifically labelled for Grover’s algorithm, with the \textit{Oracle} and \textit{Diffuser} blocks explicitly labelled. The model could simply read off the marked states directly from the \textit{Oracle} block, without needing to trace through the diffusion operator and amplitude amplification logic. This is further suggested by the format of their step-by-step reasoning traces, which solely focused on extracting the \textit{Oracle} block and identifying the marked states, without considering the rest of the circuit. It therefore remains an open question whether such an approach enhances LLMs' understanding of quantum algorithm logic, and whether it can generalise across diverse quantum algorithms or even to circuits with no recognisable algorithmic structure.

\section{Methodology}\label{sec:method}

\subsection{Datasets}
\label{sec:datasets}
We curate two datasets of quantum circuits: the \textit{Non-parameterised set} and the \textit{parameterised set}. The \textit{Non-parameterised set} consists of the following gates: H, X, Z, CNOT, CZ, multi-controlled X and Z gates with up to 4 controls. The \textit{parameterised set} is an extension of the Non-parameterised set by adding parameterised gates $R_x, R_y, \text{and } R_z$. The rotation angle $\theta$ is randomly sampled from $\{\pm\tfrac{\pi}{6}, \pm\tfrac{\pi}{4}, \pm\tfrac{\pi}{3}, \pm\tfrac{\pi}{2}\}$.

All circuits are randomly generated by sequentially sampling gates from their respective gate sets. Both datasets include circuits ranging from 1--5 qubits and 1--50 gates. For each gate, the generator first selects among single-qubit, two-qubit, or multi-qubit gates with probabilities of 75\%, 15\%, and 10\%, respectively.

All multi-qubit gates (MCX, MCMT) operate on all $n$ qubits, using the first $n-1$ as controls and the last as the target. These gates only have an effect when all control qubits are in the $\ket{1}$ state. Therefore, before applying a multi-qubit gate, the generator checks if the current state includes non-zero amplitude in the the subspace spanned by states with all control qubits in $\ket{1}$. If not, it applies Hadamard gates to all qubits with 80\% probability to introduce non-zero amplitude into the all-controls-$\ket{1}$ subspace, ensuring the dataset includes non-trivial multi-controlled samples.

All circuits are represented using the Qiskit Python API. We prefer this representation over OpenQASM primarily for token efficiency when encoding multi-controlled gates. OpenQASM does not support native multi-controlled gates; each such gate must instead be expressed through a custom gate declaration with an explicit decomposition into primitive one- and two-qubit operations, which incurs additional tokens. In contrast, Qiskit expresses the same operation in a single line, for example, \texttt{circuit.mcx([0,1,2,3], 4)} for a 4-control MCX gate. This token saving compounds across circuits that contain multi-controlled operations and reduces the length of model inputs during training and inference.

\subsection{SFT with state-vector reasoning}
\label{sec:sft-pipeline}
The SFT approach trains the model to follow a structured reasoning process that mirrors state-vector classical simulation.  The state of an n-qubit circuit is represented by a vector with $2^n$ complex numbers. One gate is applied per step and the state vector is updated by multiplying the current state by the unitary matrix representing the quantum gate. Amplitudes are represented with both symbolic expressions and floating-point numbers. We maintain a predefined list of common quantum values including common values introduced by Hadamard gates in 1--5 qubit systems ($1/\sqrt{2}$, $1/2$, $1/(2\sqrt{2})$, $1/4$, $1/(4\sqrt{2})$), and common rotation-induced values ($\sqrt{3}/2$). Amplitudes matching these values are presented in symbolic form. Values that are not in this list are represented as floating-point numbers rounded to 2 decimal places. After the step-by-step computation, we calculate the measurement probability for each basis state by taking the squared magnitude of its amplitude. Finally, the model predicts the top-15 measured states in the form of a JSON dictionary, where the key is a bitstring, and the value is the measurement probability corresponds to the bitstring rounded to 3 decimal places.  We also introduce two pairs of special tokens for the structured reasoning, \texttt{<circuit\_reasoning>}  \texttt{</circuit\_reasoning>} and \texttt{<quantum\_state>} \texttt{</quantum\_state>}. The \texttt{<circuit\_reasoning>} pair and the \texttt{<quantum\_state>} pair are used to enclose the reasoning trace and the intermediate state vector, respectively. The SFT training configuration is included in Appendix \ref{sec:sft-training-config}. Figure~\ref{fig:sft-template} and Figure~\ref{fig:example-prompt-completion}  present the overall prompt template and an example, respectively.

\begin{figure}[H]
\centering
\textbf{Reasoning trace structure}
\begin{minipage}{0.95\textwidth}
\begin{verbatim}
<circuit_reasoning>
Initialize an [n]-qubit quantum circuit in the state |[0*n]>.

[Step number]. [Python code]. [Natural Language Description].
<quantum_state>[State Vector Array]</quantum_state>

...

The probability distribution of the measurement outcome is:
  |[bitstring_1]⟩: |[amplitude_1]|² = [probability_1]
  |[bitstring_2]⟩: |[amplitude_2]|² = [probability_2]
  ...
</circuit_reasoning>
{"[bitstring_1]": [probability_1], "[bitstring_2]": [probability_2], ...}
\end{verbatim}
\end{minipage}
\caption{State-vector reasoning template for quantum circuit simulation.}
\label{fig:sft-template}
\end{figure}

\begin{figure}[H]
\centering
\textbf{Input (Prompt)}
\begin{minipage}{0.95\textwidth}
\begin{verbatim}
Simulate this quantum circuit. Then provide the probability 
distribution of the measurement outcomes as: 
{"<bitstring>": <probability>, ...}

Requirements:
- Probabilities rounded to 3 decimal places
- Sort by probability (descending)
- Only include non-zero probabilities
- Top 15 states only if >15 non-zero probabilities

Quantum Circuit: 
circuit = QuantumCircuit(1)
circuit.x(0)
circuit.h(0)
circuit.h(0)
circuit.z(0)
\end{verbatim}
\end{minipage}

\vspace{1em}
\textbf{Output (Completion)}
\begin{minipage}{\textwidth}
\small
\begin{verbatim}
<circuit_reasoning>Initialize a 1-qubit quantum circuit in the state |0⟩.

1. circuit.x(0). X gate on qubit 0.  <quantum_state>[0, 1]</quantum_state>
2. circuit.h(0). H gate on qubit 0.  <quantum_state>[1/√2, -1/√2]</quantum_state>
3. circuit.h(0). H gate on qubit 0.  <quantum_state>[0, 1]</quantum_state>
4. circuit.z(0). Z gate on qubit 0.  <quantum_state>[0, -1]</quantum_state>

The probability distribution of the measurement outcome is:
  |1⟩: |-1|² = 1.000
</circuit_reasoning>
{"1": 1.000}
\end{verbatim}
\end{minipage}
\caption{Full prompt-completion example for SFT training on a 1-qubit Non-parameterised circuit.}
\label{fig:example-prompt-completion}
\end{figure}

\subsection{Two-Stage Training with SFT and GRPO}
\label{sec:sft-grpo-pipeline}

\subsubsection{Stage 1: SFT}

\paragraph{Dataset.}
\label{sec:stage1-sft-dataset}
We restart from the base model and apply multi-task SFT rather than directly using the SFT model trained solely on the quantum circuit set, to avoid catastrophic forgetting in LLMs \cite{kirkpatrick2017overcoming}. Catastrophic forgetting refers to the complete loss of a model's previously acquired general knowledge and capabilities due to fine-tuning on highly specialised tasks. This results in \textit{entropy collapse} \cite{zhai2023stabilizing}, which refers to the loss of response diversity in model outputs and a collapse to deterministic behaviour. An example output of the Qwen3-8B model after SFT training demonstrating this issue is shown in Appendix~\ref{sec:sft-entropy-collapse-example}. This is problematic for GRPO training because the algorithm relies on comparing diverse responses to compute relative advantages for policy gradients. When the model outputs almost identical reasoning traces across rollouts, all samples receive the same rewards, resulting in zero relative advantages and no gradient signal for policy updates.

For quantum circuit reasoning, we use our own synthesised gate-by-gate simulation traces; for the remaining reasoning tasks, we draw from publicly available datasets whose reasoning traces were generated by other reasoning models, as described below. This multi-task objective preserves general reasoning ability and maintains sufficient output diversity for effective GRPO training, while still instilling the quantum circuit reasoning capability needed for Stage 2.

We construct a balanced multi-task training dataset spanning three reasoning domains, each contributing 50,000 examples for a total of 150,000 training examples:

\textbf{Quantum Circuit Reasoning.} The circuit reasoning dataset is a subset of the dataset described in Section~\ref{sec:datasets}, consisting of 50,000 randomly-generated circuits with complete gate-by-gate state-vector simulation traces as described in Section~\ref{sec:sft-pipeline}. The number of gates in the circuits is equally distributed.

\textbf{OpenMathReasoning \cite{moshkov2025aimo2}.} To maintain mathematical reasoning proficiency, we draw examples from this large-scale reasoning dataset derived from Art of Problem Solving (AoPS) forums. The original dataset comprises 3.2M CoT solutions, 1.7M tool-integrated reasoning solutions, and 566K GenSelect examples, all generated by DeepSeek-R1 \cite{guo2024deepseek} and QwQ-32B \cite{qwen2025qwen25technicalreport}. We randomly sample 50,000 examples from the dataset with two filtering criteria: (1) retain only DeepSeek-R1 CoT inference mode examples, and (2) exclude examples exceeding 17,500 tokens to maintain compatibility with the training context window.

\textbf{OpenCodeReasoning \cite{ahmad2025opencodereasoning}.} To maintain algorithmic and code reasoning capabilities, we sample from this synthetic coding dataset containing 735,255 competitive programming solutions generated by DeepSeek-R1. We apply the same 17,500-token length filtering rule and perform random sampling to obtain 50,000 examples from the filtered set.

\paragraph{Training Configuration.}

\sloppy During SFT, we mask the loss for the special token \texttt{<circuit\_reasoning>}, which prohibits the model from learning to generate the special token during inference. This training mask allows controllable reasoning modes. We can enable or disable model circuit-specific reasoning by including or excluding the \texttt{<circuit\_reasoning>} token in the prompt. As shown by Carbonneaux et al. \cite{carbonneaux2025cwm}, this approach successfully transfers knowledge to the model while preserving its ability to reason in alternative ways when the special token is absent. This property is essential for our GRPO stage, where we deliberately exclude \texttt{<circuit\_reasoning>} from prompts and instead allow the model to explore diverse reasoning strategies using its default \texttt{<think>} token. The model thus retains its knowledge of quantum circuit reasoning acquired during SFT, while also having the flexibility to discover more effective reasoning paths.

\subsubsection{Stage 2: GRPO}
\label{sec:stage2-grpo}

The second stage applies GRPO to push LLMs toward more token-efficient quantum state predictions. While SFT with simulation traces enables detailed gate-by-gate quantum circuit reasoning, it imposes a fixed reasoning strategy that may not be optimal for either accuracy or token efficiency. The model learns solely from these demonstrations without the opportunity to explore alternative reasoning approaches. GRPO \cite{shao2024deepseekmath} incentivises exploration of diverse reasoning strategies while maintaining prediction accuracy through sampling multiple candidate responses per prompt and scoring relatively via reward signals.

\paragraph{Reward Function.}
\label{sec:grpo-reward}

We design a two-level reward structure that evaluates both format compliance and probability estimation accuracy:

\textbf{Level 1: Format Validation.} The first level checks whether the model produces a valid probability distribution.

We extract the predicted distribution from the model's output by locating the JSON dictionary after the \texttt{</think>} closing tag, as Qwen3 models enclose intermediate reasoning in \texttt{<think>...</think>} tags by default during generation. The predicted distribution must satisfy all the following criteria:

\begin{enumerate}
    \item \textbf{Valid JSON formats:} Parseable as \texttt{\{"bitstring": probability, ...\}}
    \item \textbf{Correct bitstring length:} All keys are binary strings matching the circuit's qubit count
    \item \textbf{Valid probability values:} All values are numeric and within $[0, 1]$
    \item \textbf{No duplicates:} Each bitstring appears at most once
    \item \textbf{Normalised distribution:} The predicted probabilities distribution sum within $[0, 1]$
\end{enumerate}

The model receives a binary format reward:

\begin{equation}
R_{\text{format}} = \begin{cases}
  1.0 & \text{if all format criteria are satisfied}\\
  0.0 & \text{otherwise}
\end{cases}
\end{equation}

\textbf{Level 2: Total Variation Distance (TVD).} The second level evaluates the precision of probability estimates using TVD between predicted and true probability distributions. We compute TVD over all $2^n$ computational basis states:

\begin{equation}
\text{TVD} = \frac{1}{2} \sum_{i=1}^{2^n} |p_{\text{pred}}(s_i) - p_{\text{true}}(s_i)|,
\end{equation}
where $s_i$ represents the $i$-th basis state. For any state not explicitly included in the model's top-15 outputs, we assign $p_{\text{pred}}(s_i) = 0$. Similarly, for ground truth states with amplitudes below the reporting threshold ($|\alpha| < 10^{-6}$), we assign $p_{\text{true}}(s_i) = 0$.

The model receives a binary reward based on a TVD threshold $\epsilon$:

\begin{equation}
R_{\text{TVD}} = \begin{cases}
  1.0 & \text{if TVD} \leq \epsilon\\
  0.0 & \text{otherwise}
\end{cases}
\end{equation}

We calibrate the threshold $\epsilon$ separately for the non-parameterised and parameterised circuit sets. For non-parameterised circuits, we apply a stricter threshold of $\epsilon = 0.01$, as their discrete nature makes near-exact state reproduction more tractable and provides a sufficient positive training signal. For parameterised circuits, the original strict threshold yielded a negligible positive signal during early training, as continuous-valued rotation angles introduce a substantially larger solution space that the model cannot reliably explore under sparse reward conditions. We therefore adopt a relaxed threshold of $\epsilon = 0.05$ for the parameterised set, ensuring the reward gradient is maintained throughout training. 

The final reward combines format reward and TVD reward:

\begin{equation}
R_{\text{composite}} = R_{\text{format}} \cdot R_{\text{TVD}}.
\end{equation}

\paragraph{Training Configuration.}
We apply GRPO to the Qwen3-8B model using the VeRL GRPO training framework \cite{sheng2025hybridflow}. The overall GRPO training configuration is included in Appendix \ref{sec:grpo-training-config}. We use LoRA \cite{hu2022lora} with a rank=16 and alpha=32 to fine-tune the model, given the intensive computational cost of GRPO training. The model trained on Non-parameterised set and the model trained on Parameterised set share the same GRPO training configuration, except for the number of training steps and the TVD threshold $\epsilon$ used in the reward function, as described in Section~\ref{sec:grpo-reward}. The Non-parameterised set model is trained for 641 steps with $\epsilon = 0.01$, while the Parameterised set model is trained for 1,152 steps with $\epsilon = 0.05$. The difference in training steps reflects the varying convergence rates observed.

\subsection{Evaluation Metrics}
\subsubsection{General Metrics}

We evaluate model performance using a set of metrics designed to capture both prediction quality and generation efficiency. These metrics include Format Accuracy, F1-Score, TVD, Token Count, and Token Efficiency.

The token efficiency is defined as $\tau(M, D) = \frac{Q(M, D; \epsilon)}{C(M, D)}$, where $Q(M,D;\epsilon)$ measures solution quality and $C(M,D)$ denotes the token cost. The solution quality is defined as:

\[
Q(M, D; \epsilon) = \frac{1}{|D|} \sum_{i=1}^{|D|}
\mathbbm{1}\!\left[ F1_i^M = 1 \land TVD_i^M \leq \epsilon \right]
\]

To account for small deviations in the predicted probability distributions, we additionally report token efficiency under two TVD thresholds: $\epsilon \in \{0, 0.05\}$.

\subsubsection{Step-by-Step State Fidelity (Metric Specific to SFT pipeline)}

In the SFT pipeline, the model is trained on state-vector simulation traces, which serve as explicit supervision for each intermediate quantum state. To assess the quality of intermediate quantum state predictions at each gate application, we compute the quantum state fidelity between the predicted and true quantum state at each gate application step. For quantum states $\ket{\psi_{\text{pred}}}$ and $\ket{\psi_{\text{true}}}$, fidelity is defined as the inner product of the two states, denoted as $F = |\langle \psi_{\text{true}} | \psi_{\text{pred}} \rangle|^2$, ranging from 0 to 1. Unlike classical probability distribution metrics that operate solely on real-valued probabilities, quantum fidelity properly accounts for the complex-valued amplitudes $\alpha_i \in \mathbb{C}$ and their relative phases. This metric evaluates whether the model correctly tracks quantum states in the complex Hilbert space $\mathbb{C}^{2^n}$. 

\subsection{Evaluation Setup}
We compare the performance of our two training pipelines against two baselines: the base model Qwen3-8B \cite{yang2025qwen3} and one of the state-of-the-art open-source reasoning models, GPT-OSS-120B \cite{agarwal2025gpt}. Table~\ref{tab:eval-sets} summarises the three evaluation sets we used to evaluate model performances.

\begin{table}[h]
\centering
\caption{Evaluation sets configurations.}
\label{tab:eval-sets}
\begin{tabular}{llcc}
\toprule
\textbf{Set} & \textbf{Type} & \textbf{Qubits} & \textbf{Gates} \\
\midrule
Set 1 & In-Distribution         & 1--5 & 1--50  \\
Set 2 & Gate Count Extrapolation & 1--5 & 51--70 \\
Set 3 & System Size Extrapolation & 6--7 & 1--20  \\
\bottomrule
\end{tabular}
\end{table}

All models are evaluated using the same ground-truth distributions, but GPT-OSS-120B is evaluated under a different qubit ordering convention. GPT-OSS-120B recognises the Qiskit code representation that Qiskit uses Least Significant Bit (LSB) ordering and reasoned accordingly. All other models (Qwen3-8B, SFT, and SFT+GRPO) reason using Most Significant Bit (MSB) ordering, where qubit 0 is the leftmost bit. This MSB convention is used consistently in our SFT training data, as it aligns with the standard mathematical notation for quantum state vectors and is the default convention adopted by Qwen3-8B without explicit prompting. To ensure a fair comparison, ground-truth distributions are converted to match each model's inferred ordering convention before computing evaluation metrics. Illustrative excerpts from GPT-OSS-120B and Qwen3-8B reasoning traces confirming the two conventions are provided in Appendix~\ref{sec:ordering-appendix}.

\subsection{Inference Configuration}
All models are evaluated with a temperature of 1.0 and top-p of 0.95. The maximum generation length is set to 32,768 tokens to allow more budget for longer reasoning traces, especially for circuits in extrapolation sets 2 and 3.

\section{Results}\label{sec:results}
\subsection{Overall Results}
Table \ref{tab:comprehensive-results} summarises the performance across evaluation Sets 1--3 in terms of F1-score and TVD. SFT achieves the highest F1-scores on Sets 1 and 2 for both circuit sets and the lowest TVD across both sets, indicating accurate prediction of measurement probabilities. However, SFT completely fails to generalise to larger system sizes on Set 3, with F1 of 0.000 and TVD of 1.000 on the Non-parameterised set.

SFT+GRPO achieves F1-scores close to those of SFT on Sets 1 and 2, but with higher TVD. This indicates that SFT+GRPO correctly predicts the top-15 computational basis elements, albeit with lower precision in the predicted probability distribution. Crucially, SFT+GRPO achieves non-zero F1-scores on Set 3, demonstrating that the GRPO stage allows the LLM to explore reasoning strategies that generalise to larger state spaces.

Both fine-tuned pipelines significantly outperform the baselines. Qwen3-8B achieves F1-scores of only 0.616 and 0.724 on Non-parameterised and Parameterised Set 1, respectively, well below both fine-tuned models. GPT-OSS-120B, despite being a much larger model, achieves an F1-score of 0.729 on Non-parameterised Set 1 but collapses to 0.293 on Parameterised Set 1, reflecting its difficulty with the larger output space introduced by continuous rotation angles. Despite the substantially larger solution space introduced by continuous rotation angles, both SFT and SFT+GRPO show only a modest decline in F1-scores from Non-parameterised to Parameterised circuits on Sets 1 and 2, suggesting that the structured reasoning approach generalises well across both circuit sets.

\begin{table}[h]
\centering
\caption{Top-15 F1-Score and TVD across evaluation sets for non-parameterised and parameterised circuits. Set~1 is the in-distribution test set, Set~2 evaluates gate-count extrapolation, and Set~3 evaluates system-size extrapolation. Bold entries indicate the best result per column. SFT and SFT+GRPO significantly outperform the baselines on Set 1 and Set 2, while SFT+GRPO provides stronger generalisation to larger qubit systems than SFT alone.}
\label{tab:comprehensive-results}

\begin{tabular}{lccccccc}
\toprule
\multirow{3}{*}{\textbf{Model}} & \multicolumn{3}{c}{\textbf{Non-parameterised Set}} & \multicolumn{3}{c}{\textbf{Parameterised Set}} \\
\cmidrule(lr){2-4} \cmidrule(lr){5-7}
& \textbf{Set 1} & \textbf{Set 2} & \textbf{Set 3} & \textbf{Set 1} & \textbf{Set 2} & \textbf{Set 3} \\
\midrule
\multicolumn{7}{l}{\textbf{Top-15 F1-Score}} \\
SFT & \textbf{0.988} & \textbf{0.948} & 0.000 & \textbf{0.952} & \textbf{0.919} & 0.084 \\
SFT+GRPO & 0.967 & 0.900 & 0.045 & 0.918 & 0.902 & 0.265 \\
\midrule
Qwen3-8B & 0.616 & 0.523 & 0.476 & 0.724 & 0.607 & 0.428 \\
GPT-OSS & 0.729 & 0.400 & \textbf{0.721} & 0.293 & 0.028 & \textbf{0.549} \\
\midrule
\multicolumn{7}{l}{\textbf{Top-15 TVD}} \\
SFT & \textbf{0.016} & \textbf{0.068} & 1.000 & \textbf{0.030} & \textbf{0.058} & 0.917 \\
SFT+GRPO & 0.045 & 0.149 & 0.955 & 0.149 & 0.381 & 0.751 \\
\midrule
Qwen3-8B & 0.448 & 0.595 & 0.532 & 0.491 & 0.638 & 0.587 \\
GPT-OSS & 0.282 & 0.617 & \textbf{0.290} & 0.718 & 0.982 & \textbf{0.433} \\
\hline
\end{tabular}%

\end{table}

\subsection{SFT training pipeline}
\subsubsection{Training and Validation Loss}
\begin{figure}[H]
    \centering
    \begin{subfigure}{0.48\textwidth}
        \centering
        \includegraphics[width=\textwidth]{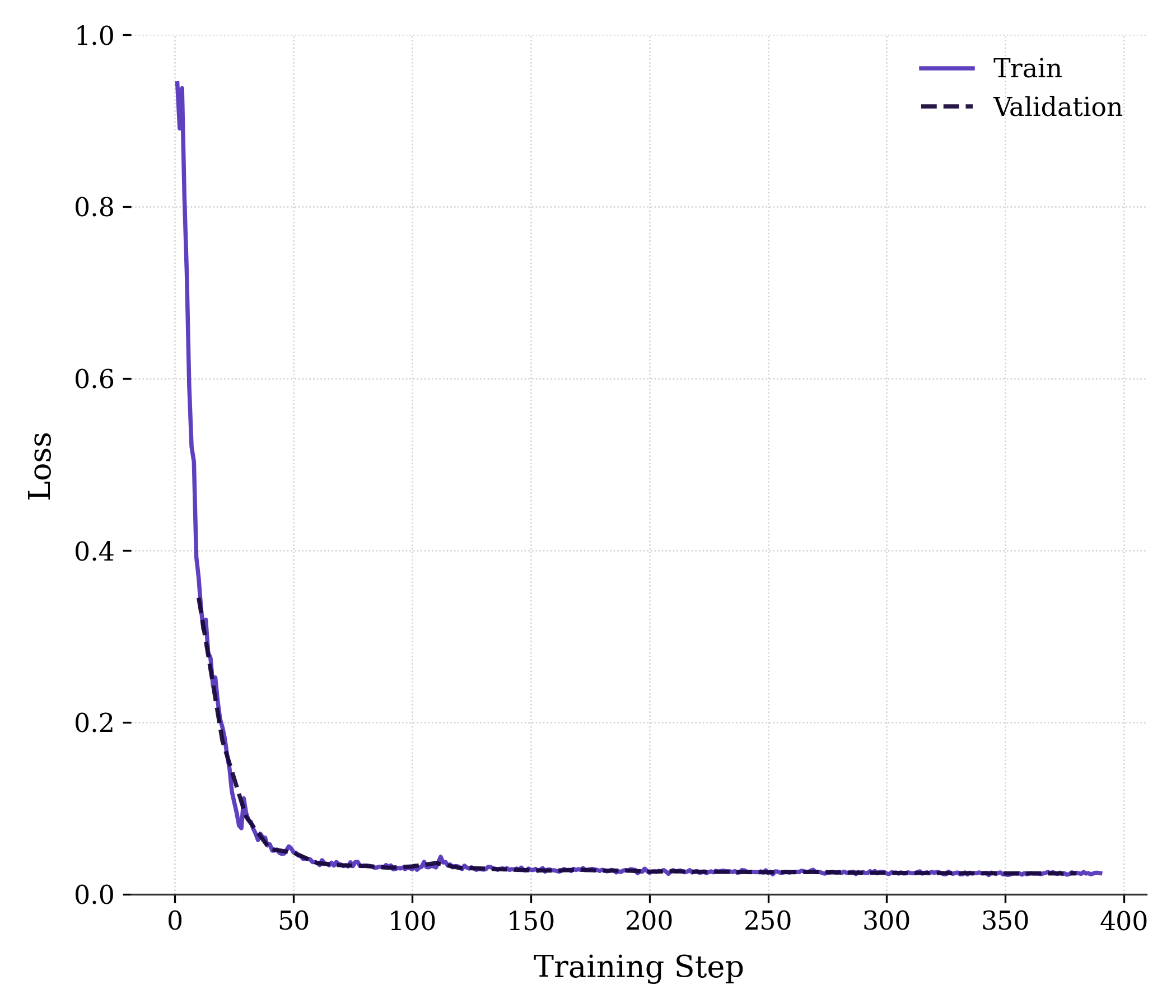}
        \caption{Non-parameterised Set}
        \label{fig:sft_train_val_loss_non_parameterised}
    \end{subfigure}
    \hfill
    \begin{subfigure}{0.48\textwidth}
        \centering
        \includegraphics[width=\textwidth]{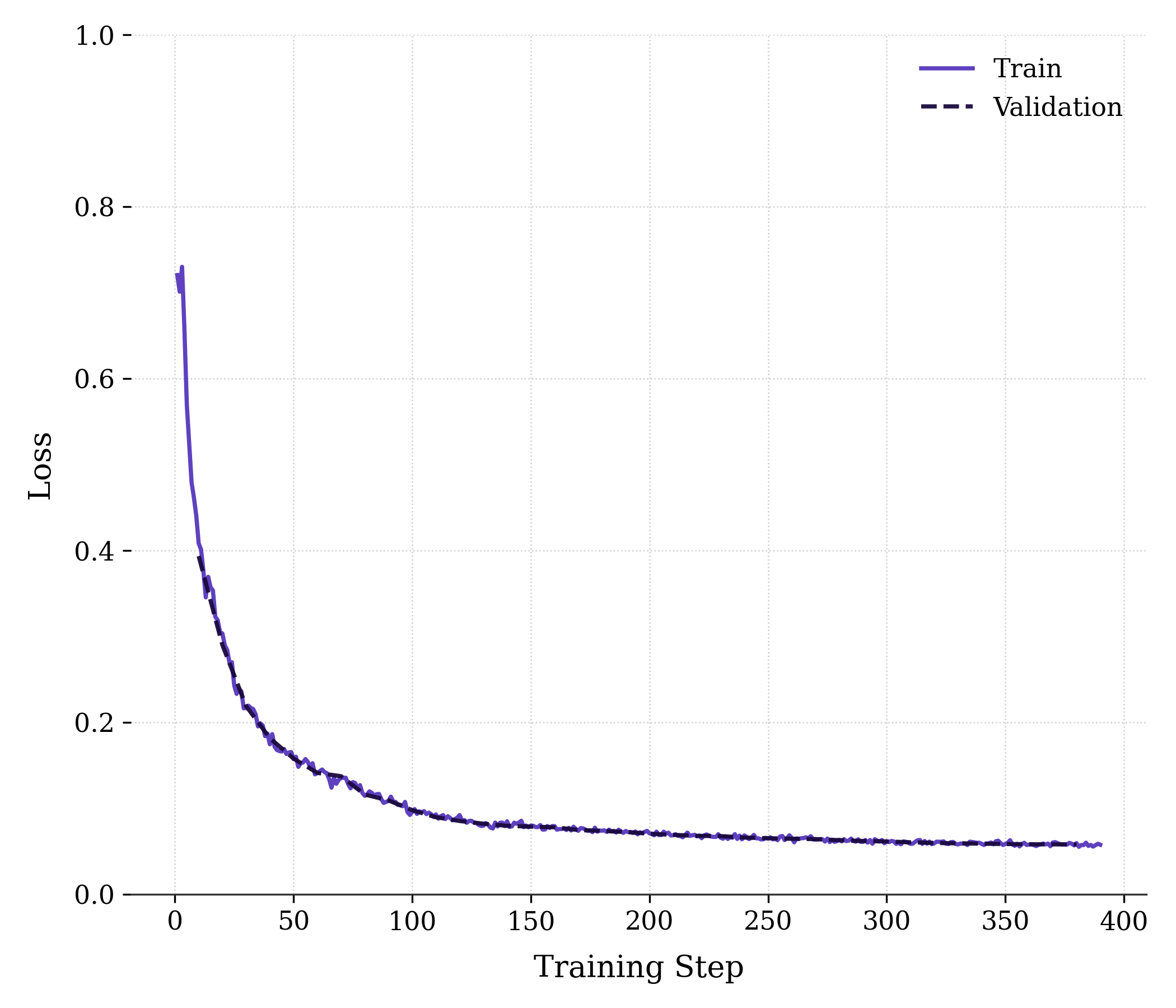}
        \caption{Parameterised Set}
        \label{fig:sft_train_val_loss_parameterised}
    \end{subfigure}
    \caption{Train and validation loss of SFT training for the Non-parameterised (left) and Parameterised (right) circuit sets. Both sets converge within one epoch (390 steps).}
    \label{fig:sft_train_val_loss}
\end{figure}

Figure \ref{fig:sft_train_val_loss_non_parameterised} and Figure \ref{fig:sft_train_val_loss_parameterised} show the training dynamics on the Non-parameterised set and Parameterised set, respectively. There are a total of 390 steps for one training epoch and the training on both sets converged within one epoch.

\subsubsection{Step-by-Step State Fidelity}
\begin{figure}[H]
    \centering
    \begin{subfigure}{\textwidth}
        \centering
        \includegraphics[width=\textwidth]{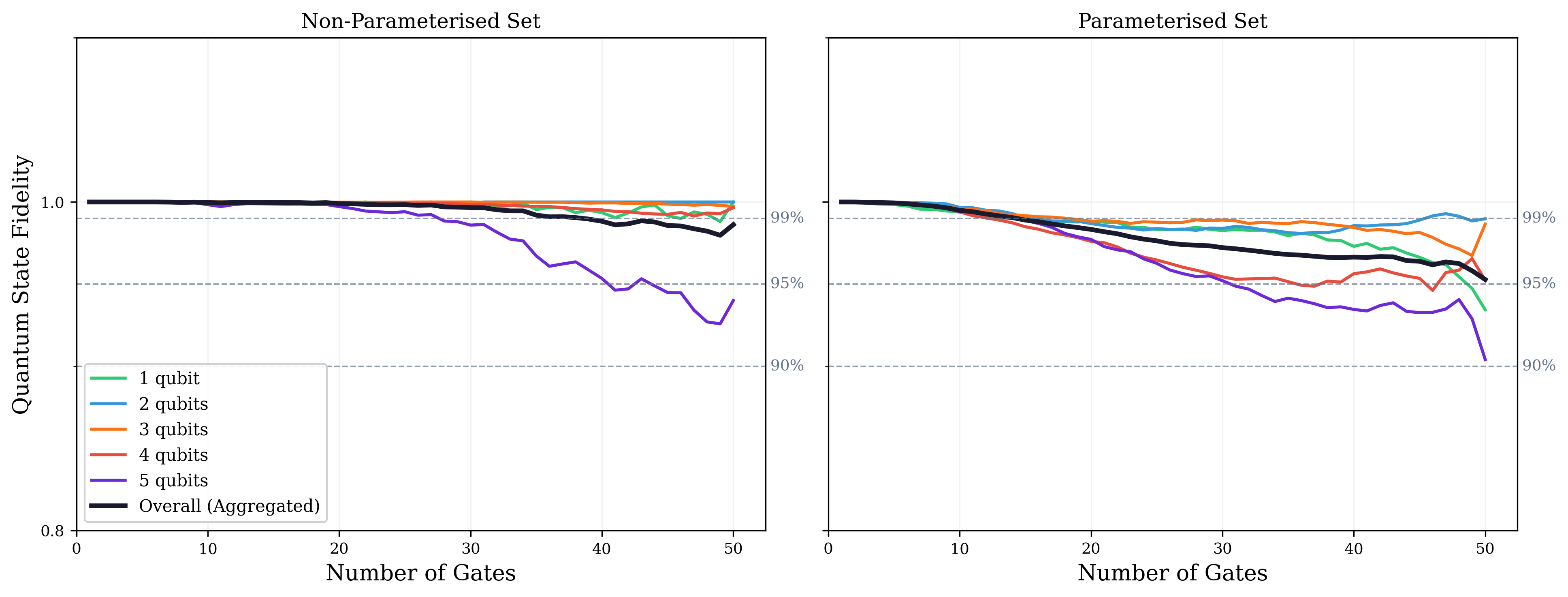}
    \end{subfigure}
    \caption{Step-by-step quantum state fidelity of the SFT model during inference, evaluated on the Non-parameterised (left) and Parameterised (right) circuit sets. Fidelity remains above 99\% throughout for the Non-parameterised set, while the Parameterised set exhibits a steeper decline with gate depth yet stays above 95\% in aggregate, demonstrating that the model maintains a coherent intermediate quantum state representation across both settings. Notably, 5-qubit circuits show the most pronounced fidelity decline in both sets, reflecting the increased difficulty of tracking quantum state evolution over circuits with greater system size. }
    \label{fig:step_by_step}
\end{figure}

In the SFT pipeline, LLMs are trained to generate the resulting quantum state after each gate application. We compute the quantum state fidelity with respect to the ground truth at each intermediate step to assess whether the trained model maintains a coherent quantum state representation throughout the sequence of gates. Figure \ref{fig:step_by_step} shows the step-by-step fidelity for the Non-parameterised set and Parameterised set. 

For the Non-parameterised set, quantum state fidelity remains consistently high across all circuit sizes, with the overall aggregated fidelity remaining above 99\%. Individual qubit configurations exhibit similarly stable trajectories, with fidelity degrading only marginally with additional gates. The 5-qubit circuits show the greatest decline, yet remain above 95\% throughout, indicating that the model maintains a coherent quantum state representation even for more complex non-parameterised circuits.

The Parameterised set presents a more challenging setting, where fidelity declines more sharply with increasing gate count. The 5-qubit circuits again show the steepest degradation, with fidelity approaching the 95\% threshold at larger gate counts. Nevertheless, the overall aggregated fidelity remains above 95\% across both circuit sets, demonstrating that the SFT model can maintain reliable intermediate quantum state representations throughout gate-by-gate reasoning.

\subsubsection{Ablation Study}
\begin{table}[H]
\centering
\caption{Ablation comparison across Non-parameterised and Parameterised sets. The SFT model trained with intermediate reasoning traces significantly outperforms the SFT model trained without traces, confirming that step-by-step reasoning provides critical learning signal for quantum circuit reasoning.}
\label{tab:ablation-random-sft}

\begin{tabular}{lcccccc}
\toprule
\multirow{2}{*}{\textbf{Model}} & \multicolumn{3}{c}{\textbf{Non-parameterised Set}} & \multicolumn{3}{c}{\textbf{Parameterised Set}} \\
\cmidrule(lr){2-4} \cmidrule(lr){5-7}
& \textbf{Set 1} & \textbf{Set 2} & \textbf{Set 3} & \textbf{Set 1} & \textbf{Set 2} & \textbf{Set 3} \\
\midrule
\multicolumn{7}{l}{\textbf{Top-15 F1-Score}} \\
SFT w/ reasoning  & \textbf{0.988} & \textbf{0.948} & 0.000 & \textbf{0.952} & \textbf{0.919} & \textbf{0.084} \\
SFT w/o reasoning & 0.571 & 0.534 & 0.000 & 0.895 & 0.916 & 0.000 \\
\midrule
\multicolumn{7}{l}{\textbf{Top-15 TVD}} \\
SFT w/ reasoning  & \textbf{0.016} & \textbf{0.068} & 1.000 & \textbf{0.030} & \textbf{0.058} & \textbf{0.917} \\
SFT w/o reasoning & 0.493 & 0.572 & 1.000 & 0.409 & 0.483 & 1.000 \\
\bottomrule
\end{tabular}%
\end{table}

To isolate the effect of intermediate reasoning traces on model performance, we retrained the base model on the same set of circuits without the ground-truth reasoning traces. The models are trained with SFT under the same conditions as listed in Appendix~\ref{sec:sft-training-config}. 

The results in Table~\ref{tab:ablation-random-sft} show that the SFT model trained without reasoning traces performs substantially worse than the full SFT model, as performance on both F1-score and TVD metrics degrades significantly across all evaluation sets. This indicates that the step-by-step reasoning provides critical learning signals that enable the model to acquire accurate quantum circuit simulation skills. However, the ablation SFT model still achieves moderate performance on in-distribution Set 1 and gate-count extrapolation Set 2, suggesting that the model can learn some aspects of quantum transformations from circuit structure or pattern alone, even without intermediate reasoning traces. Both models completely fail on Non-parameterised Set 3. This failure appears to stem from the length generalisation bottleneck discussed in Section~\ref{sec:qualitative} and Section~\ref{sec:discussion}, rather than from missing reasoning traces, since it affects both models equally regardless of the training signal.

\subsection{SFT+GRPO training pipeline}
\begin{figure}[H]
\centering
\includegraphics[width=1.0\textwidth]{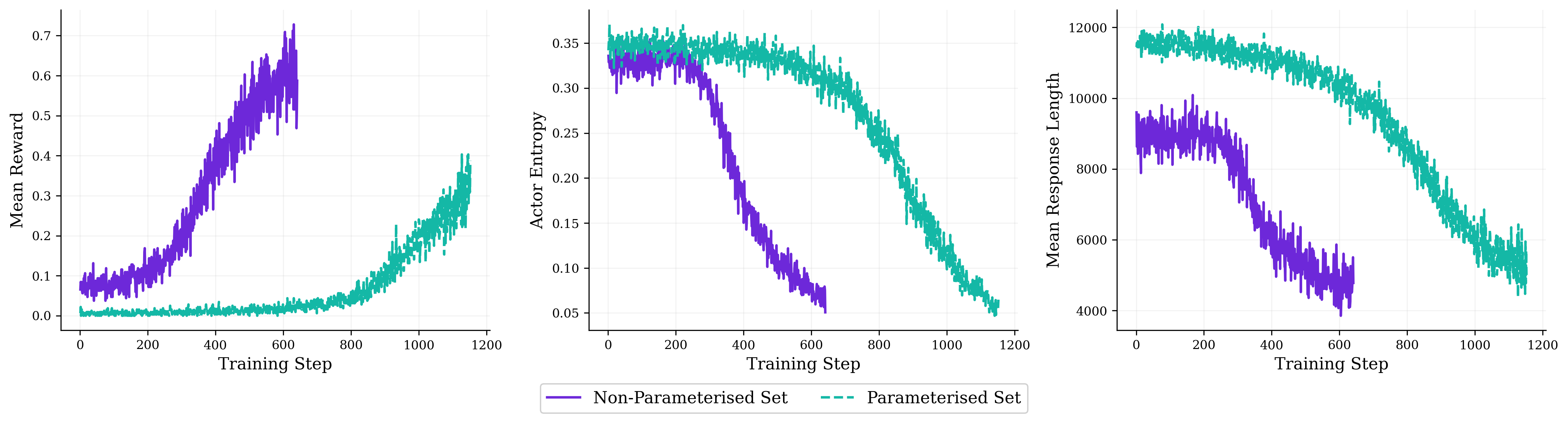}
\caption{Training dynamics of the GRPO stage for the Non-parameterised and Parameterised circuit sets. Mean reward (left), actor entropy (centre), and mean response length (right) are plotted against training step. Both sets show a pattern of rising reward that is accompanied by declining entropy and shorter responses, indicating that the model converges toward more accurate and concise reasoning strategies. The Non-parameterised set converges faster than Parameterised set, reflecting the greater task complexity of parameterised circuits. Despite the difference in reward levels, entropy and response length converge to similar values across both sets by the end of training.}
\label{fig:grpo_training_metrics}
\end{figure}

Figure \ref{fig:grpo_training_metrics} compares the mean reward, actor entropy, and mean response length during GRPO training for the Non-parameterised and Parameterised circuit sets. Across both sets, the increase in mean reward over training steps correlates with a decrease in entropy and mean response length, indicating that the model learns to generate more accurate and concise reasoning traces as training progresses. The Non-parameterised set exhibits a more rapid increase in mean reward and a sharper decline in entropy and response length compared to the Parameterised set, consistent with the lower task complexity of non-parameterised circuits. The Non-parameterised set reached a mean reward of 0.7 after 641 steps, while the Parameterised set reached only 0.3 after 1,152 steps, reflecting the different convergence rates and reward thresholds applied to each set. Despite the difference in mean reward, the model entropy and mean response length for both sets converge to similar values by the end of training.

\subsubsection{GRPO Contribution and Token Budget Effect}
\label{sec:grpo-contribution}

\begin{figure}[H]
\centering
\includegraphics[width=0.9\textwidth]{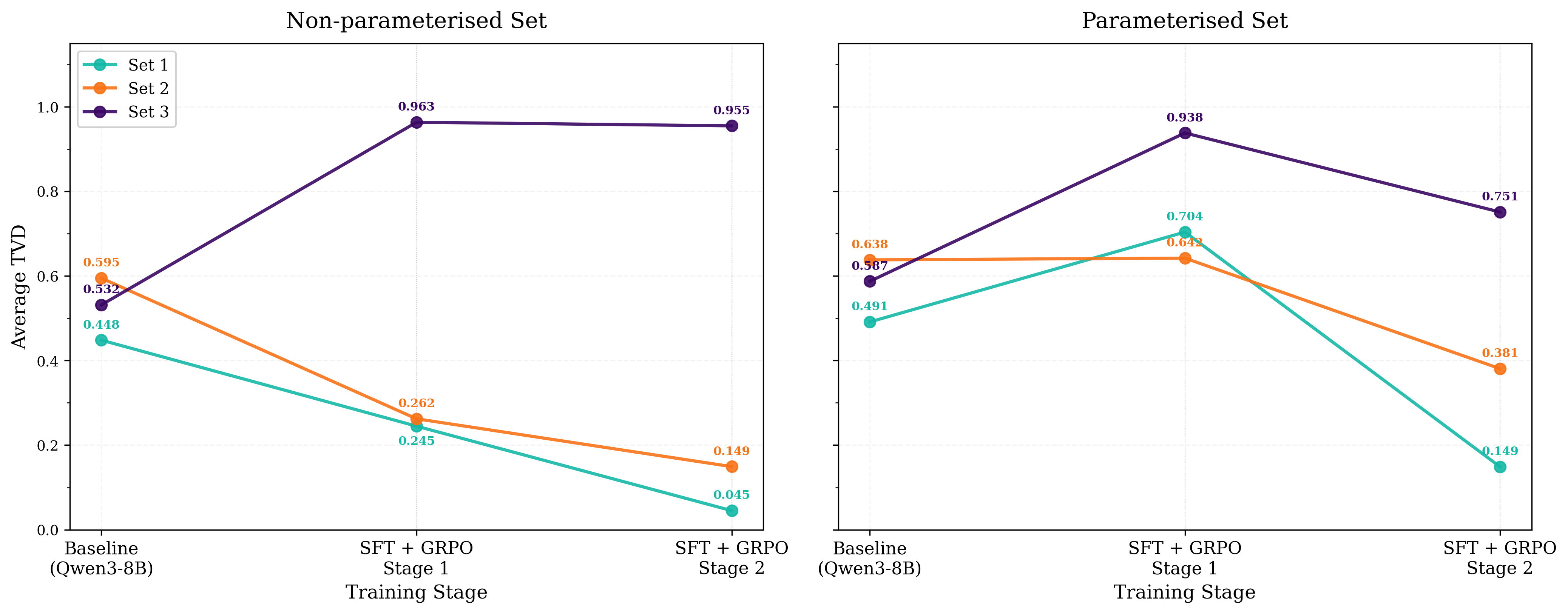}
\vspace{0.5em}
\includegraphics[width=0.9\textwidth]{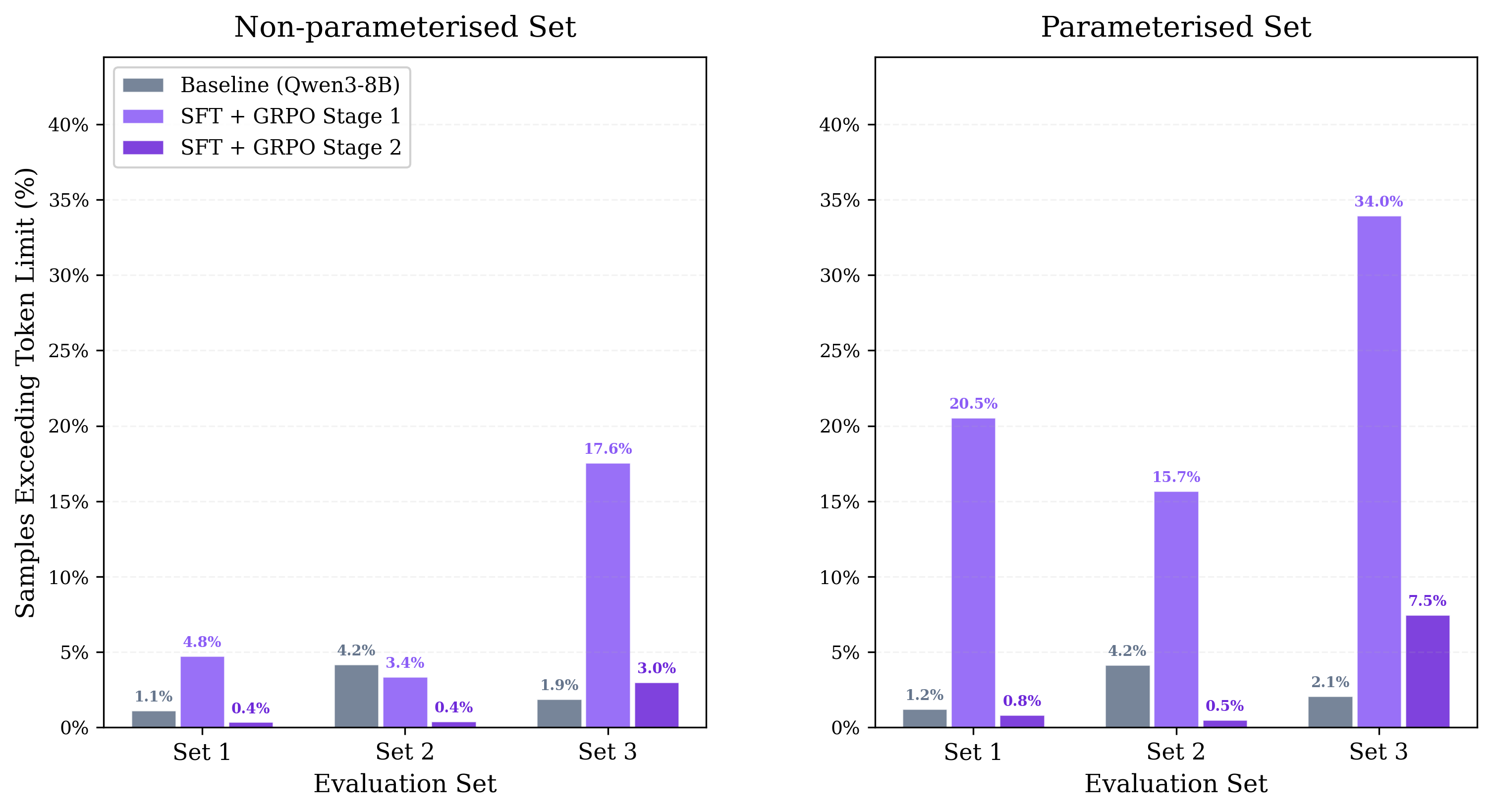}
\caption{Three-stage TVD progression and token-limit violation counts. (top) Mean TVD across the base model, Stage 1 SFT, and SFT+GRPO for the Non-parameterised and Parameterised sets on the three evaluation sets. (bottom) Number of samples whose generation exceeds the 32,768-token limit at each stage. The sharp rise in token-limit violations at Stage 1 SFT for the Parameterised set explains the elevated TVD at that stage. Stage 2 GRPO suppresses this by learning more concise reasoning strategies.}
\label{fig:stage1-grpo-contribution}
\end{figure}

To isolate the contribution of the GRPO stage, we evaluate the intermediate Stage 1 SFT model alongside the base model and the final SFT+GRPO model. Figure~\ref{fig:stage1-grpo-contribution} summarises this three-stage TVD progression together with the number of samples that exceed the token generation limit at each stage.

For the Non-parameterised set, Stage 1 SFT already achieves a significant reduction in TVD over the base model, and Stage 2 GRPO provides a further improvement. The pattern differs substantially for the Parameterised set, where Stage 1 SFT yields a TVD that is higher than or comparable to the base model, before Stage 2 GRPO brings it down significantly. The token-limit violation count explains this anomaly. The multi-task SFT dataset includes long-form reasoning traces from mathematical and code competition problems, which causes the Stage 1 SFT model to generate substantially longer outputs when faced with parameterised quantum circuits. As a result, a considerably larger fraction of parameterised samples exhaust the 32,768-token generation limit at Stage 1 SFT than at either the base model or after Stage 2 GRPO. Truncated outputs cannot produce a well-formed final probability distribution, directly inflating TVD. This effect is far less significant on the Non-parameterised set, whose circuits require shorter state-vector traces and therefore produce fewer token-limit violations.

The decrease in mean response length after Stage 2 GRPO shows that the model learned to generate more concise reasoning strategies through Stage 2 GRPO. This conciseness reduces token-limit violations on the Parameterised set, recovering prediction quality and achieving the TVD improvement seen in the final SFT+GRPO results.

\subsection{Performance Scaling}
\begin{figure}[H]
\centering
\includegraphics[width=0.9\textwidth]{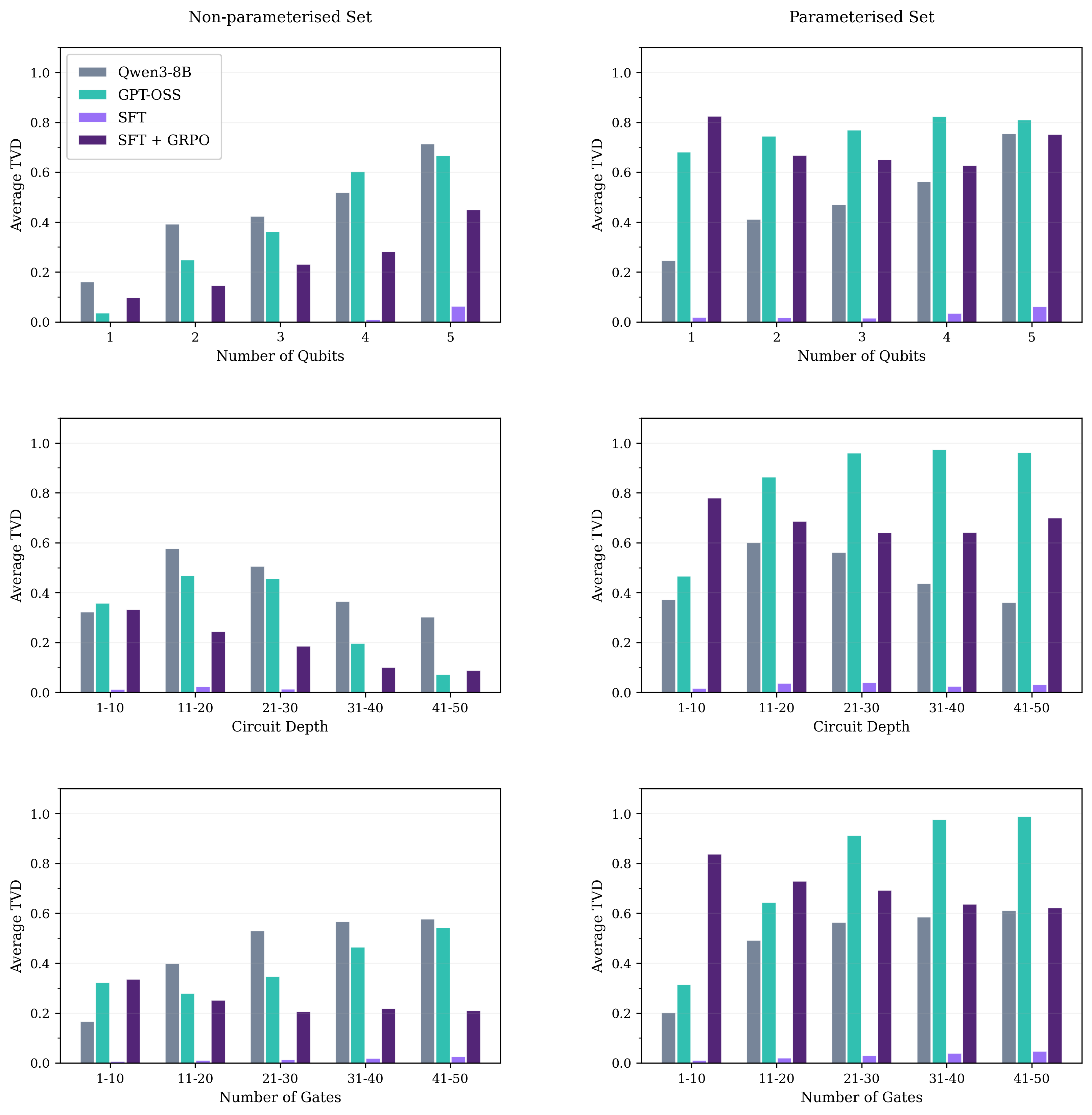}
\caption{Mean TVD as a function of circuit complexity for the Non-Parameterised Set (left column) and Parameterised Set (right column), broken down by number of qubits (top), circuit depth (middle), and number of gates (bottom). The number of qubits shows the strongest correlation with TVD degradation, followed by the number of gates, while circuit depth has the least influence.}
\label{fig:performance-scaling}
\end{figure}

Figure \ref{fig:performance-scaling} shows how model performance, measured by average TVD, scales with three circuit complexity dimensions. Among the three dimensions, the number of qubits shows the strongest correlation with performance degradation, the number of gates has a moderate effect, and circuit depth has the least influence. These findings align with our step-by-step reasoning framework. Since the model processes each gate sequentially, tracking the quantum state through individual operations, an increase in circuit depth does not add complexity beyond the required sequential reasoning steps. Therefore, circuit depth does not correlate with TVD decline. Conversely, the number of gates increases the problem linearly, with each additional gate introducing an extra reasoning step and increasing the potential for error accumulation. Meanwhile, the quantum state space grows exponentially with the number of qubits, placing greater demands on the model’s capacity for accurate state tracking and directly driving the observed increase in TVD.

\subsection{Token Efficiency}
Table~\ref{tab:token-efficiency} reports average token counts and token efficiency across all evaluation sets. The result shows that SFT achieves the highest token efficiency on in-distribution and gate-count extrapolation tasks, driven by both strong accuracy and low token cost. On Non-parameterised Set 1, SFT achieves an efficiency of 0.424 while generating an average of 2,101 tokens, which is substantially fewer than SFT+GRPO's 2,609. The structured gate-by-gate reasoning format instilled during training produces concise outputs with high prediction accuracy, and token usage scales predictably with circuit complexity, growing from 2,101 on Set 1 to 4,896 on Set 2.

On the system-size extrapolation set (Set 3), SFT's efficiency collapses on non-parameterised circuits but not entirely on parameterised circuits. For the Non-parameterised set, SFT produces short outputs (1,902 tokens) but with zero correct predictions, causing $\tau$ to collapse to zero regardless of token cost. For the Parameterised set, however, SFT achieves a small but non-zero efficiency of 0.017. SFT+GRPO achieves non-zero accuracy on the Non-parameterised set that SFT cannot, and the higher token counts generated by SFT+GRPO on this set (3,697 and 5,816 respectively) reflect the reasoning effort over a larger state space. Notably, GPT-OSS averages 32,218 and 32,455 tokens on Parameterised Sets 1 and 2, which are both close to the 32,768-token generation limit, suggesting the model is attempting to perform long reasoning traces that exceed the context window, which may contribute to its poor performance on these sets.

\begin{table}[H]
\centering
\caption{Token count and efficiency metrics across evaluation sets. Average Token Count reports the mean number of tokens generated per response. Token Efficiency is defined as the ratio of correctly simulated samples (TVD $<$ 0.05 or TVD $=$ 0) to total tokens used, measuring how accurately a model performs per token spent. Fine-tuned models (SFT, SFT+GRPO) use substantially fewer tokens than baselines while achieving far higher accuracy, indicating significantly better token efficiency. Bold values indicate the best result per column.}
\label{tab:token-efficiency}
\begin{tabular}{lcccccc}
\toprule
\multirow{3}{*}{\textbf{Model}} & \multicolumn{3}{c}{\textbf{Non-parameterised Set}} & \multicolumn{3}{c}{\textbf{Parameterised Set}} \\
\cmidrule(lr){2-4} \cmidrule(lr){5-7}
& \textbf{Set 1} & \textbf{Set 2} & \textbf{Set 3} & \textbf{Set 1} & \textbf{Set 2} & \textbf{Set 3} \\
\midrule
\multicolumn{7}{l}{\textbf{Average Token Count}} \\
SFT       & 2101 & 4896  & 1902 & 3924  & 10308 & 2522  \\
SFT+GRPO  & 2609 & 5171  & 3697 & 4449  & 8832  & 5816  \\
\midrule
Qwen3-8B  & 10811 & 11033  & 9904 & 11879 & 9019  & 10682 \\
GPT-OSS   & 15095 & 26106 & 9293 & 32218 & 32455 & 14725 \\
\midrule
\multicolumn{7}{l}{\textbf{Token Efficiency (TVD $<$ 0.05)}} \\
SFT       & \textbf{0.424} & \textbf{0.159} & 0.000 & \textbf{0.125} & \textbf{0.021} & 0.017 \\
SFT+GRPO  & 0.328          & 0.133          & 0.005 & 0.106          & 0.018          & 0.013 \\
\midrule
Qwen3-8B  & 0.032          & 0.013          & 0.033 & 0.013          & 0.002          & 0.021 \\
GPT-OSS   & 0.041          & 0.013          & \textbf{0.055} & 0.008          & 0.000          & \textbf{0.022} \\
\midrule
\multicolumn{7}{l}{\textbf{Token Efficiency (TVD $=$ 0)}} \\
SFT       & \textbf{0.316} & \textbf{0.097} & 0.000 & \textbf{0.016} & 0.000          & 0.008 \\
SFT+GRPO  & 0.249          & 0.085          & 0.003 & 0.012          & 0.000          & 0.007 \\
\midrule
Qwen3-8B  & 0.030          & 0.012          & 0.030 & 0.004          & 0.000          & \textbf{0.010} \\
GPT-OSS   & 0.036          & 0.011          & \textbf{0.047} & 0.002          & 0.000          & 0.008 \\
\bottomrule
\end{tabular}%
\end{table}

\subsection{Qualitative Analysis}
\label{sec:qualitative}

We now conduct a qualitative analysis of model outputs to understand the reasoning patterns and failure modes exhibited by each training pipeline.

\subsubsection{Token-efficient shortcuts for SFT+GRPO pipeline}
\label{sec:qualitative-grpo}

Figure~\ref{fig:sft-like-reasoning-shortcut} shows a representative SFT+GRPO output on a 5-qubit circuit. A reasoning behavior emerges that is absent from the training data. When the model encounters a gate that leaves the quantum state unchanged it substitutes the natural-language phrase ``\textit{So the state remains the same.}'' rather than re-listing the full 32-entry state vector. This shortcut is never shown in the SFT training data, where every step unconditionally writes out the complete state vector. The model autonomously discovers this behavior during GRPO exploration, reducing token expenditure without sacrificing information content.

\begin{figure}[H]
\centering
\textbf{Input (Prompt, circuit only)}
\begin{minipage}{0.95\textwidth}
\small
\begin{verbatim}
circuit = QuantumCircuit(5)
circuit.h(4)
circuit.cz(0, 4)
circuit.ry(-pi/3, 0)
circuit.h(1)
circuit.h(0)
circuit.h(1)
circuit.h(2)
circuit.h(3)
circuit.h(4)
circuit.append(MCXGate(4), [0, 1, 2, 3, 4])
\end{verbatim}
\end{minipage}

\vspace{1em}
\textbf{Model reasoning trace (\texttt{<think>} block)}
\begin{minipage}{0.95\textwidth}
\small
\begin{verbatim}
<think>
Okay, let's see. I need to simulate this quantum circuit step by step
and find the probability distribution of the measurement outcomes.
The circuit is 5-qubit, and the operations are a bit complex.
Let me try to break it down.

First, the circuit starts with 5 qubits in the state
[1, 0, 0, 0, 0, 0, 0, 0, 0, 0, 0, 0, 0, 0, 0, 0,
 0, 0, 0, 0, 0, 0, 0, 0, 0, 0, 0, 0, 0, 0, 0, 0].

The first operation is circuit.h(4). This adds a 1/sqrt(2) to the
4th qubit. So the state becomes
[1/sqrt(2), 1/sqrt(2), 0, 0, 0, 0, 0, 0, 0, 0, 0, 0, 0, 0, 0, 0,
 0, 0, 0, 0, 0, 0, 0, 0, 0, 0, 0, 0, 0, 0, 0, 0].

Next, circuit.cz(0, 4). CZ gate on qubits [0,4], which doesn't
change the state because both are 0. So the state remains the same.

Then, circuit.ry(-pi/3, 0). RY(-pi/3) gate on qubit 0. The state
becomes
[0.61, 0.61, 0, 0, 0, 0, 0, 0, 0, 0, 0, 0, 0, 0, 0, 0,
 -0.35, -0.35, 0, 0, 0, 0, 0, 0, 0, 0, 0, 0, 0, 0, 0, 0].

\end{verbatim}
\end{minipage}
\caption{Example SFT+GRPO output illustrating the token-efficient shortcut behavior absent from the SFT training demonstrations. After \texttt{circuit.h(4)} places qubit~4 into superposition, the subsequent \texttt{CZ(0,\,4)} gate has no effect because the control qubit~0 remains in $\ket{0}$. Rather than re-listing the full 32-entry state vector, the model writes ``\textit{So the state remains the same.}''}
\label{fig:sft-like-reasoning-shortcut}
\end{figure}

\subsubsection{System-Size Extrapolation}

Figure~\ref{fig:system-size-extrapolation} shows outputs from both pipelines on the same 6-qubit circuit, which lies outside the 5-qubit training distribution. Both models share a common failure in the intermediate reasoning trace: the state vector is initialized with 32 entries rather than the correct 64 entries and gate applications proceed throughout in this reduced 5-qubit space. Neither model successfully tracks all 64 basis states during step-by-step simulation.

Despite this shared intermediate failure, the two pipelines diverge in their final answers. The SFT+GRPO model produces 6-character bitstrings in its probability distribution, correctly matching the 6-qubit circuit size, even though its internal reasoning operated in 5-qubit space. The SFT model, by contrast, outputs 5-character bitstrings, propagating the reduced dimensionality all the way to the final prediction. This contrast illustrates the partial length generalisation advantage of the GRPO stage.

\begin{figure}[H]
\centering
\textbf{Circuit (shared prompt)}
\begin{minipage}{0.95\textwidth}
\small
\begin{verbatim}
circuit = QuantumCircuit(6)
circuit.x(3)
circuit.h(0)
circuit.h(1)
\end{verbatim}
\end{minipage}

\vspace{1em}
\begin{minipage}[t]{0.47\textwidth}
\textbf{SFT+GRPO output}
\noindent\begin{lstlisting}[breaklines=true, basicstyle=\ttfamily\scriptsize, breakatwhitespace=false, columns=flexible, xleftmargin=0pt, breakindent=0pt, literate={√}{$\surd$}1]
<think>
Okay, let's see. I need to simulate a quantum circuit and find the probability distribution after some operations. The circuit starts with 6 qubits, all in the state [1, 0, 0, 0, 0, 0, 0, 0, 0, 0, 0, 0, 0, 0, 0, 0, 0, 0, 0, 0, 0, 0, 0, 0, 0, 0, 0, 0, 0, 0, 0, 0].

The first operation is circuit.x(3). X gate on qubit 3. So the state becomes [0, 0, 1, 0, 0, 0, 0, 0, 0, 0, 0, 0, 0, 0, 0, 0, 0, 0, 0, 0, 0, 0, 0, 0, 0, 0, 0, 0, 0, 0, 0, 0].

Next, circuit.h(0). H gate on qubit 0. The state becomes [0, 0, 1/√2, 0, 0, 0, 0, 0, 0, 0, 0, 0, 0, 0, 0, 0, 0, 0, 1/√2, 0, 0, 0, 0, 0, 0, 0, 0, 0, 0, 0, 0, 0].

Then, circuit.h(1). H gate on qubit 1. The state becomes [0, 0, 1/2, 0, 0, 0, 0, 0, 0, 0, 1/2, 0, 0, 0, 0, 0, 0, 0, 1/2, 0, 0, 0, 0, 0, 0, 0, 1/2, 0, 0, 0, 0, 0].

So the probability distribution is:
|000100>: |1/2|^2 = 0.250
|010100>: |1/2|^2 = 0.250
|100100>: |1/2|^2 = 0.250
|110100>: |1/2|^2 = 0.250

Each of these four states has a probability of 0.25. So the output should be these four states with 0.25 each.
</think>{"000100": 0.25, "100100": 0.25, "010100": 0.25, "110100": 0.25}
\end{lstlisting}
\end{minipage}
\hfill
\begin{minipage}[t]{0.47\textwidth}
\textbf{SFT output}
\noindent\begin{lstlisting}[breaklines=true, basicstyle=\ttfamily\scriptsize, breakatwhitespace=false, columns=flexible, xleftmargin=0pt, breakindent=0pt, literate={√}{$\surd$}1]
<circuit_reasoning>Initialize a 6-qubit quantum circuit in the state [1, 0, 0, 0, 0, 0, 0, 0, 0, 0, 0, 0, 0, 0, 0, 0, 0, 0, 0, 0, 0, 0, 0, 0, 0, 0, 0, 0, 0, 0, 0, 0].
1. circuit.x(3). X gate on qubit 3. <quantum_state>[0, 0, 1, 0, 0, 0, 0, 0, 0, 0, 0, 0, 0, 0, 0, 0, 0, 0, 0, 0, 0, 0, 0, 0, 0, 0, 0, 0, 0, 0, 0, 0]</quantum_state>

2. circuit.h(0). H gate on qubit 0. <quantum_state>[0, 0, 1/√2, 0, 0, 0, 0, 0, 0, 0, 0, 0, 0, 0, 0, 0, 0, 0, 1/√2, 0, 0, 0, 0, 0, 0, 0, 0, 0, 0, 0, 0, 0]</quantum_state>

3. circuit.h(1). H gate on qubit 1. <quantum_state>[0, 0, 1/2, 0, 0, 0, 0, 0, 0, 0, 1/2, 0, 0, 0, 0, 0, 0, 0, 1/2, 0, 0, 0, 0, 0, 0, 0, 1/2, 0, 0, 0, 0, 0]</quantum_state>

The probability distribution of measurement outcome is:
|00100>: |1/2|^2 = 0.250
|01100>: |1/2|^2 = 0.250
|10100>: |1/2|^2 = 0.250
|11100>: |1/2|^2 = 0.250
</circuit_reasoning>{"00100": 0.25, "10100": 0.25, "01100": 0.25, "11100": 0.25}
\end{lstlisting}
\end{minipage}
\caption{System-size extrapolation on a 6-qubit circuit. Both models initialise and maintain a 32-entry state vector throughout the reasoning trace instead of the correct 64 entries, failing to track the full 6-qubit state space. However, SFT+GRPO recovers at the output stage and produces 6-character bitstrings, whereas SFT carries the reduced dimensionality into the final answer and outputs 5-character bitstrings.}
\label{fig:system-size-extrapolation}
\end{figure}

\section{Discussion and Conclusion}\label{sec:discussion}
The ability to reason about quantum state evolution is central to quantum algorithm development and verification. In this work, we investigate whether LLMs can be trained to perform structured quantum reasoning. We fine-tune Qwen3-8B on curated datasets of non-parameterised and parameterised circuits spanning 1–5 qubits and 1–50 gates. We propose two fine-tuning pipelines: (i) SFT on gate-by-gate state-vector simulation traces, and (ii) a two-stage SFT+GRPO pipeline that instils structured reasoning via SFT, then applies GRPO with verifiable rewards to encourage exploration of more token-efficient reasoning strategies. Both pipelines significantly outperform the base model and GPT-OSS-120B. This shows that targeted fine-tuning on structured reasoning traces is more effective than zero-shot prompting, even for much larger models. SFT achieves the strongest in-distribution accuracy and token efficiency, and generalise well to gate-count extrapolation set. However, it collapses entirely on qubit counts beyond the training distribution. SFT+GRPO achieves comparable in-distribution F1 and does not collapse entirely on qubit-count extrapolation. This is enabled by adaptive reasoning strategies elicited by the GRPO stage. An ablation study further shows that models can approximate quantum dynamics from circuit descriptions without explicit step-by-step reasoning traces. This suggests that the gate sequence itself encodes information that LLMs can use to approximate quantum state evolution, though with reduced numerical precision compared to reasoning with simulation traces.

Furthermore, models fine-tuned with both pipelines handle parameterised circuits well despite the substantially larger solution space introduced by the continuous rotation angles, resulting in only a modest performance decline relative to the non-parameterised setting. This suggests that the structured reasoning approach generalises well across circuit types, and that the added complexity of continuous parameters does not fundamentally alter the nature of the task for the model. The more substantial challenge arises not from circuit parameterisation but from length generalisation.

Length generalisation---the ability to handle sequences longer than those seen during training---has long been identified as a fundamental challenge for LLMs, particularly in tasks requiring arithmetic or algorithmic reasoning~\cite{anil2022exploring, jelassi2023length, zhou2023algorithms}. Despite ongoing research, it remains an open problem. Our results exhibit a similar pattern: while SFT+GRPO demonstrates partial length generalisation by correctly producing output bitstrings with the appropriate length for larger qubit systems, it still struggles to maintain accurate state tracking over the correspondingly larger state space during step-by-step reasoning. This highlights that the exponential scaling of quantum state representations with system size is a concrete instance of the broader length generalisation challenge, one that targeted fine-tuning alone is unlikely to fully resolve.

Beyond the length generalisation challenge discussed above, we identify two further limitations in our experimental design. First, our simulation traces use limited numerical precision---amplitudes are rounded to two decimal places as a deliberate trade-off for token efficiency---which imposes a precision ceiling on learned representations and makes it difficult to disentangle genuine reasoning failures from cumulative rounding errors. Future work could address this by designing richer symbolic representations that extend beyond substituting a fixed set of common quantum values. Second, following prior quantum simulation work~\cite{wang2024grovergpt,chen2026symbolic}, our evaluation targets probability distributions only, discarding all phase information from quantum state vectors; while SFT's intermediate reasoning steps permit some assessment of full state-vector understanding, this offers limited insight into the quantum reasoning capabilities of the SFT+GRPO pipeline.

For future directions, having demonstrated that gate-by-gate simulation traces provide a strong training signal for quantum reasoning, a natural next step is to extend this paradigm toward a \textit{Quantum World Model}: a model that internalises quantum knowledge and applies this grounded understanding to downstream tasks such as circuit synthesis and algorithm design, rather than relying on superficial pattern matching from training data. Additionally, this work represents circuits using the Qiskit Python API; extending the framework to alternative representations such as OpenQASM 3.0 or Cirq would broaden its applicability and enable direct comparison with benchmarks that adopt different circuit description languages.

\bibliographystyle{unsrt}
\bibliography{sn-bibliography}

\newpage
\appendix
\section{SFT Training Configuration} 
\label{sec:sft-training-config}

\begin{table}[h]
\centering
\begin{tabular}{ll}
\toprule
\textbf{Configuration} & \textbf{Value} \\
\midrule
Base model & Qwen3-8B \\
Maximum sequence length & 17,500 tokens \\
Batch size & 256 \\
Learning rate & $1 \times 10^{-4}$ \\
Weight decay & 0.01 \\
Optimizer & AdamW ($\beta_1 = 0.9$, $\beta_2 = 0.95$) \\
LR scheduler & Cosine with 10\% warmup \\
Training epochs & 1 \\
Hardware & 4 H100 GPUs (1 nodes $\times$ 4) \\
\bottomrule
\end{tabular}
\end{table}

\section{Example Output from SFT Model with a Random Prompt}
\label{sec:sft-entropy-collapse-example}

\begin{figure}[H]

\textit{Input: ``How are you?''} \\

\textit{Output:} \\
Initialize a 3-qubit quantum circuit in the state $\ket{000}$.

\medskip
\begin{enumerate}
  \item \lstinline[style=circuitstyle]{circuit.rz(-3*pi/4, 2)}. 
        $\mathrm{RZ}(-3\pi/4)$ gate on qubit 2. \\
        \texttt{<quantum\_state>} $[0.38 + 0.92i,\ 0,\ 0,\ 0,\ 0,\ 0,\ 0,\ 0]$
        \texttt{</quantum\_state>}

  \item \lstinline[style=circuitstyle]{circuit.ry(-pi/4, 1)}.
        $\mathrm{RY}(-\pi/4)$ gate on qubit 1. \\
        \texttt{<quantum\_state>} $[0.35 + 0.85i,\ 0,\ {-0.15 - 0.35i},\ 0,\ 0,\ 0,\ 0,\ 0]$
        \texttt{</quantum\_state>}

  \item \lstinline[style=circuitstyle]{circuit.ry(3*pi/4, 0)}.
        $\mathrm{RY}(3\pi/4)$ gate on qubit 0. \\
        \texttt{<quantum\_state>} $[0.14 + 0.33i,\ 0,\ {-0.06 - 0.14i},\ 0,$
        $0.33 + 0.79i,\ 0,\ {-0.14 - 0.33i},\ 0]$
        \texttt{</quantum\_state>}

\end{enumerate}

\label{fig:sft-entropy-example}
\end{figure}

\section{GRPO Training Configuration}
\label{sec:grpo-training-config}

\begin{table}[h]
\centering
\begin{tabular}{ll}
\toprule
\textbf{Configuration} & \textbf{Value} \\
\midrule
LoRA rank & 16 \\
LoRA alpha & 32 \\
LoRA Target modules & All linear layers \\
Training batch size & 64 \\
Rollouts per sample & 5 \\
Rollouts Temperature & 1.0 \\
Rollouts Top-p & 1.0 \\
PPO updates per iteration & 4 \\
Learning rate & $1 \times 10^{-6}$ \\
Hardware & 32 H100 GPUs (8 nodes $\times$ 4) \\
\bottomrule
\end{tabular}
\end{table}

\section{Qubit Ordering Convention in Model Reasoning Traces}
\label{sec:ordering-appendix}
\subsection*{GPT-OSS-120B: Qiskit's LSB ordering}

The following is extracted from GPT-OSS-120B's internal reasoning when simulating a 3-qubit Qiskit circuit:
\\

\begin{minipage}{0.95\textwidth}
\begin{verbatim}
We need to consider that the circuit is defined in Qiskit, which uses
little-endian ordering for measurement bits: qubit 0 is the least
significant bit in the bitstring. [...] Qiskit uses little-endian: the
first qubit is the least significant bit. [...] for probability
distribution, we can just output bitstrings as standard binary
representation with qubit 0 as the rightmost bit.

Thus we need to compute probabilities for each basis state.
[...]

Qiskit uses little-endian ordering for state vectors:
index = q0 + 2*q1 + 4*q2. So the vector ordering is:

  0: |000> (q2=0, q1=0, q0=0)
  1: |001> (q2=0, q1=0, q0=1)
  2: |010> (q2=0, q1=1, q0=0)
  3: |011> (q2=0, q1=1, q0=1)
  4: |100> (q2=1, q1=0, q0=0)
  
[...]

\end{verbatim}
\end{minipage}

\subsection*{Qwen3-8B: Implicit MSB ordering}

The following is extracted from Qwen3-8B's reasoning when simulating the same circuit structure. The model adopts MSB ordering throughout without acknowledging the Qiskit convention:
\\

\begin{minipage}{0.95\textwidth}
\small
\begin{verbatim}
Let me denote the state as a vector of 8 elements, where each element
corresponds to the amplitude of |q0 q1 q2>.

[...]

First, initial state is |000⟩.

Let me list all the gates in order:

1. h(1) → apply H to qubit 1. State becomes: 1/√2 (|000⟩ + |010⟩)
2. h(0) → apply H to qubit 0. Now the state is 1/2 (|000⟩ + |010⟩ + |100⟩ + |110⟩)

[...]

\end{verbatim}
\end{minipage}
\end{document}